\RequirePackage{fix-cm}
\documentclass[smallextended]{svjour3}       
\smartqed  
\usepackage{multirow}
\usepackage{amsmath}

\usepackage{CJKutf8}
\usepackage{tcolorbox}
\usepackage{pgfkeys}

\usepackage{graphicx}  
\usepackage{float}  
\usepackage{subfigure}  

\usepackage{siunitx}
\usepackage{booktabs}    
\usepackage{graphicx}
\usepackage{makecell}
\usepackage{multirow}
\usepackage{longtable}
\usepackage{diagbox}
\usepackage{float}
\usepackage{cite}
\usepackage{enumerate}
\usepackage{hyperref}
\usepackage{tablefootnote} 
\usepackage{array}
\usepackage{enumitem}
\usepackage{framed}
\usepackage[linewidth=1pt]{mdframed}
\usepackage{indentfirst}
\usepackage{color}

\definecolor{hzcolor}{RGB}{10, 186, 181}

\usepackage{etoolbox}

\usepackage{bm} 

\usepackage{siunitx} 
\newrobustcmd\B{\DeclareFontSeriesDefault[rm]{bf}{b}\bfseries}
\begin{document}

\title{How to get better embeddings with code pre-trained models? An empirical study}

\author{Yu Zhao \and
        Lina Gong \and
        Haoxiang Zhang \and
        Yaoshen Yu \and
        Zhiqiu Huang
}

\institute{
            Yu Zhao  \and Lina Gong \and Yaoshen Yu \and Zhiqiu Huang \at College of Computer Science and Technology, Nanjing University of Aeronautics and Astronautics, and State Key Lab. for Novel Software Technology,  Nanjing University, and Key Laboratory of Safety-Critical Software, Nanjing University of Aeronautics and Astronautics. \\  
            \email{zhao\_yu@nuaa.edu.cn, linagong@nuaa.edu.cn, yaoshen.yu@outlook.com, zqhuang@nuaa.edu.cn}\\
            \and Haoxiang Zhang  \at
              Software Analysis and Intelligence Lab (SAIL), Queen's University, Kingston, ON, Canada \\
              \email{haoxiang.zhang@acm.org}\\
             \and Corresponding author: Zhiqiu Huang and Lina Gong, Email: zqhuang@nuaa.edu.cn, gonglina@nuaa.edu.cn\\
}
\date{Received: date / Accepted: date}

\maketitle

\begin{abstract}

Pre-trained language models have demonstrated powerful capabilities in the field of natural language processing (NLP). Recently, code pre-trained model (PTM), which draw from the experiences of the NLP field, have also achieved state-of-the-art results in many software engineering (SE) downstream tasks. These code PTMs take into account the differences between programming languages and natural languages during pre-training and make adjustments to pre-training tasks and input data. 
However, researchers in the SE community still inherit habits from the NLP field when using these code PTMs to generate embeddings for SE downstream classification tasks, such as generating semantic embeddings for code snippets through special tokens and inputting code and text information in the same way as pre-training the PTMs. 
In this paper, we empirically study five different PTMs (i.e. CodeBERT, CodeT5, PLBART, CodeGPT and CodeGen) with three different architectures (i.e. encoder-only, decoder-only and encoder-decoder) on four SE downstream classification tasks (i.e. code vulnerability detection, code clone detection, just-in-time defect prediction and function docstring mismatch detection) with respect to the two aforementioned aspects.
Our experimental results indicate that (1) regardless of the architecture of the code PTMs used, embeddings obtained through special tokens do not sufficiently aggregate the semantic information of the entire code snippet; (2) the quality of code embeddings obtained by combing code data and text data in the same way as pre-training the PTMs is poor and cannot guarantee richer semantic information; (3) using the method that aggregates the vector representations of all code tokens, the decoder-only PTMs can obtain code embeddings with semantics as rich as or even better quality than those obtained from the encoder-only and encoder-decoder PTMs.
Based on our findings, we recommend that researchers in the SE community (1) pay attention to the vector representation of each code token when generating code embeddings for downstream classification tasks, for example, obtaining richer semantic embeddings through simple average-pooling of all code tokens; (2) adopt the unimodal input approach when generating code embeddings for downstream classification tasks with both code and text information to obtain competitive and higher quality code embeddings; and (3) use larger-scale decoder architecture PTMs to aggregate vector representations of all input code tokens to obtain higher quality code embeddings with richer implied semantic information.
Our research provides SE researchers with guidance on how to obtain higher-quality code embeddings when using code PTMs, thereby advancing future research on SE downstream classification tasks.

\end{abstract}

\section{Introduction}
\label{sec:Introduction}

Distributed representations of code have played a crucial role in harnessing the power of deep learning for software engineering tasks \cite{siow-distributed-learning, kanade-distributed-learning}. These methods aim to learn low-dimensional vector representations \cite{hoang-distributed-cc2vec, alon-distributed-code2vec, hellendoorn-distributed-global}, known as code embeddings, to capture the essence of source code. Within these embeddings, the meaning of the code is distributed across multiple vector components, allowing semantically similar code snippets to be mapped to closely positioned vectors. This property empowers code embeddings to perform exceptionally well in a variety of downstream software engineering tasks, including program comprehension tasks like vulnerability detection \cite{cheng-distributed-vd}, as well as program generation tasks like code comment generation \cite{hu-distributed-cg}.

Recently, pre-trained language models (PTM) have gained immense popularity in the field of natural language processing (NLP) \cite{dale-llm, scao-llm, raffel-llm, touvron-llm}. These pre-trained models are machine learning models that are pre-trained on vast amounts of training data to learn general features and possess powerful representational capabilities \cite{feng-embedding, reimers-embedding}. They can generate sentence embeddings that encapsulate rich semantic information, which can be used for various downstream NLP tasks \cite{sun-nlp, khot-nlp, zeng-nlp}. In order to advance research in the domain of code, researchers in the Software Engineering (SE) community have developed various code PTMs based on the similarity between programming languages and natural languages. For instance, following the release of the natural language PTM BERT \cite{kenton-bert}, SE researchers subsequently introduced several PTMs capable of learning programming languages. These include CuBERT \cite{kanade-cubert}, CodeBERT \cite{feng-codebert}, TreeBERT \cite{jiang-treebert}, GraphCodeBERT \cite{guo-graphcodebert}, and more. These code PTMs can generate code embeddings that convey abundant semantic information for tasks such as programming comprehension and generation \cite{ni-pc, wei-pg}.

We know that SE researchers have taken into account the specific characteristics of code when training code PTMs, leading to the design of novel pre-training tasks to better accommodate these characteristics. For example, Feng et al. \cite{feng-codebert} took into account code-specific annotation information when training the CodeBERT model and designed a bimodal input method of code and annotation to input data and trained PL-NL pairs on the Masked Language Modeling task during training. Wang et al.\cite{wang-codet5}, on the other hand, focused on identifiers in code when training CodeT5, tailoring specific training tasks for them. However, current researchers often still follow the habit of using language models to obtain embeddings in the NLP field when using code PTMs to generate embeddings for code snippets \cite{liu-special, niu-special}. Sharma et al. \cite{sharma-attention-bert} retrained BERT on Java code and found that, in clone detection tasks, the retrained BERT paid more attention to syntactic entities, especially identifiers and delimiters in Java code, compared to the widely focused [CLS] token in NLP. Inspired by their work, we realized that despite previous studies in the SE field indicating similarities between programming languages and natural languages, it may not be entirely appropriate for current researchers to directly apply NLP knowledge to assume similar behavior in code. This is especially crucial when generating code embeddings using PTMs as the embedding is the first critical step in leveraging deep learning techniques to solve SE downstream tasks. Using an inappropriate approach may result in lower-quality code embeddings, which may fail to capture the rich semantic information in the code and subsequently impact the performance of downstream tasks.

To this end, this paper focuses on the classification task in the SE field to study the practices of researchers in the SE community when using code PTMs to generate embedding representations. Based on previous literature surveys, we found:

(1) Currently, when many researchers use code PTMs to generate embedding representations for code snippets on SE classification tasks, they often follow the practice of NLP researchers in using certain special tokens to obtain semantic aggregation embeddings of code snippets \cite{chai-special, zeng-special}. The NLP PTMs can use the special token to extract a semantic embedding representing its meaning from a piece of natural language and then convert it into an equivalent natural language description. This approach has been successful in obtaining semantic representations of sentences in the NLP domain, as it can capture semantic relationships between words in a sentence. But for a piece of code of classification tasks, the semantic embedding aggregated through such a special token may not be rich enough to capture the different information of code between different categories.

(2) For code-related classification tasks with textual information such as natural language annotations, SE researchers do not particularly focus on how code and text are combined when inputting code and text information as token sequences into those PTMs \cite{kanade-cubert, zhou-codebert4jit}. This can impact the richness of semantic information encoded in the resulting code embeddings. Many researchers, when using code PTMs to generate embeddings for code snippets containing comments, employ the method of combining code and comments as they were during pre-training. For instance, recent studies \cite{zhou-codebert4jit} utilizing CodeBERT separate code and comments using special tokens as they were during pre-training, and then pair them as input into the CodeBERT to obtain code embeddings.

In order to obtain higher quality code embeddings when using code PTMs and provide guidance to researchers in the SE community, this study explores the above two aspects using a total of five PTMs (i.e. CodeBERT \cite{feng-codebert}, CodeT5 \cite{wang-codet5}, PLBART \cite{ahmad-plbart}, CodeGPT \cite{lu-codexglue-codegpt} and CodeGen \cite{nijkamp-codegen}) from three different architectures (i.e. encoder-only, decoder-only and encoder-decoder) across four SE classification tasks (i.e. code vulnerability detection, code clone detection, just-in-time defect prediction and function docstring mismatch detection) using three performance evaluated metrics (i.e. Accuracy, F1 and MCC). Specifically, we formulate the following two research questions to guide our investigation:

\begin{itemize}
    \item \textbf{RQ1: Can the embedding obtained through a particular token sufficiently aggregate the semantic information of the entire code snippets?}
    \item \textbf{RQ2: How do the way code and text are combined affect the quality of semantic embeddings generated by code pre-trained models?}
\end{itemize}

Our experimental results demonstrate:

\begin{itemize}
    \item[$\bullet$] No matter which architecture of the code PTM is used, the embedding obtained through a special token cannot fully aggregate the semantic information of the entire code snippets. Conversely, focusing on the vector representations of each code token, for instance, through a simple average pooling method, leads to embeddings with richer semantic information. This approach benefits the PTMs in encoder-only architectures to a lesser extent, followed by encoder-decoder architectures, and provides the most benefit to models in decoder-only architectures.
    \item[$\bullet$] No matter which architecture of the code PTM is used, the quality of code embedding obtained by inputting data according to the way of combining code and text information during pre-training is generally poor, and it is not guaranteed to obtain code embedding with richer semantic information. In contrast, inputting code information and text information separately i.e. unimodal input proves to be a competitive approach for obtaining higher-quality code embeddings.
\end{itemize}

\textbf{Significance of Study.} Our study provides new insights into the use of code PTMs by researchers in the current SE community to generate code embeddings. This includes the current SE field's thinking on the use of PTms borrowed from the NLP field and how to use these code PTMs to exert their capabilities to generate code embeddings with richer semantic information to support downstream classification tasks. SE researchers can leverage the findings presented in this paper to effectively employ current code PTMs and potentially achieve improved results. We have open-sourced our experimental code and data, facilitating replication, result validation, and further dissemination within the SE community.

\textbf{Paper Organization.} Section 2 discusses the background and motivation for this study. Section 3 introduces the experimental setup of the empirical study and provides an overview of our study. Section 4 describes the two research questions of our work, including research methods,  results analysis and discussion. Section 5 shows the implications of the experimental conclusions. Section 6 discusses threats to the validity of our experimental conclusions. Finally, Section 7 concludes the paper.

\section{Motivation and Related work}
\label{sec:Motivation}

Pre-trained language models have shown great promise in the NLP field. BERT, as a representative example, has achieved advanced performance in various downstream NLP tasks. Drawing parallels between programming languages and natural languages, researchers in the SE community have explored code PTMs based on their studies of BERT. CuBERT \cite{kanade-cubert} was the first to propose a programming language-based PTM for learning code embedding representations. It employs the same architecture and pre-training tasks as BERT to model the Python language for obtaining code embeddings. Unlike natural language, programming language possesses unique semantic information. To generate distributed vector representations for code, CodeBERT \cite{feng-codebert} employs a structure similar to BERT, but with different training tasks, modeling six programming languages to obtain a unified code representation. In contrast to natural language, the structural information of code in the programming language is more crucial. Therefore, some researchers have incorporated code structure information into the training data to obtain universal code PTMs such as TreeBERT \cite{jiang-treebert} and GraphCodeBERT \cite{guo-graphcodebert}. These PTMs have demonstrated superior performance in downstream tasks related to program comprehension and program generation. For example, the CodeBERT proposed by Feng et al. \cite{feng-codebert} has shown good performance in code search tasks, while the CodeT5 proposed by Wang et al. \cite{wang-codet5} has performed well in tasks like code translation, defect detection, and code summarization. Nijkamp et al.'s CodeGen \cite{nijkamp-codegen} has also shown strong performance in code generation tasks.

In addition to directly applying PTMs to specific downstream tasks, many researchers also utilize these PTMs to generate code embeddings for further integration into the training pipeline of task-specific models. Zhou et al. \cite{zhou-codebert4jit} extracted information from the code embeddings generated by CodeBERT using convolutional neural networks and achieved performance comparable to the current state-of-the-art method in the just-in-time defect prediction task. Sun et al. \cite{sun-bert-smart}utilized the BERT model to obtain feature representations of smart contract code, and combined active learning techniques with uncertain sampling strategies to learn information related to contract vulnerabilities from these feature representations, achieving good performance in contract vulnerability detection. Tang et al. \cite{tang-csgvd} proposed the CSGVD method, which combines BiLSTM with embeddings generated by CodeBERT to effectively detect vulnerabilities based on code embeddings.

Ding et al. \cite{ding-can} extended the experiments conducted by Kang et al. \cite{kang-assess} on six downstream SE tasks, including code comment generation, code authorship identification, code clone detection, source code classification, log statement prediction, and software defect prediction. They found that using code embedding techniques indeed contributed to achieving better performance in SE downstream tasks. However, we observed that researchers often acquire the embeddings generated for code classification tasks in a manner similar to how NLP researchers use special tokens to obtain semantic embeddings for text snippets. For example, when Feng et al. \cite{feng-codebert} released the CodeBERT model, they recommended and used a special token, [CLS], representing the vector at the first position as the semantic embedding for the entire input. Similarly, when Lu et al. \cite{lu-codexglue-codegpt} released the CodeGPT model, they used a special token, [SEP], representing the vector at the last position as the semantic embedding for the entire input. Subsequent researchers often followed this approach to obtain code embeddings. This method of obtaining embeddings using special tokens stems from the NLP field and may not be suitable for the SE domain, potentially leading to lower-quality embeddings \cite{liu-special, niu-special}. Therefore, we pose the first research question:

RQ1: Can the embedding obtained through a particular token sufficiently aggregate the semantic information of the entire code snippets?

Furthermore, we observe that many researchers have not paid particular attention to how code snippets with accompanying text information are combined \cite{liu-ccrep, zhou-codebert4jit}. Different methods of combining them as input to PTMs can also impact the quality of the resulting code embeddings. Therefore, we raise the second research question:

RQ2: How do the way code and text are combined affect the quality of semantic embeddings generated by code pre-trained models?

Code embedding is a crucial step in downstream tasks, as high-quality code embeddings can encapsulate rich semantic information and facilitate the practicality of subsequent tasks. To address the aforementioned research issues, we conducted a study on how to obtain higher-quality code embeddings across four SE classification tasks using a total of five pre-trained models with three different architectures. The next section will provide a detailed overview of the techniques involved.

\section{Experimental Design}
\label{sec: design}

In this section, we provide a detailed description of the key components involved in the experimental design of this paper. This includes the PTMs utilized for generating code embeddings, the downstream SE classification tasks employed for experimental research, as well as the associated task datasets and the metrics used for performance evaluation.

\subsection{Pre-trained Models}
To ensure the comprehensiveness of the experiments in this paper and the applicability of the conclusions, we examine code PTMs across all three different architectures: encoder-only, encoder-decoder, and decoder-only. Specifically, we select CodeBERT \cite{feng-codebert} for the encoder-only architecture, and for the decoder-only architecture, we employ CodeGPT \cite{lu-codexglue-codegpt} and CodeGen \cite{nijkamp-codegen}. In the case of the encoder-decoder architecture, our choices are CodeT5 \cite{wang-codet5} and PLBART \cite{ahmad-plbart}. Note that these five code PTMs are all pre-trained on multiple programming languages and the pre-training data includes both code data and text data. These code-related PTMs are widely utilized by researchers in the recent SE community for various downstream tasks. A brief introduction to these five models is provided in Table \ref{tab:ptms}.

\begin{table*}[h]
\centering
  \caption{Five code PTMs of three architectures for generating embeddings}
  \begin{tabular}{cccc}
    \toprule
    Name & Architecture & Parameter Size  & Embedding Dimension \\
    \midrule
   CodeBERT &Encoder         &125M  &768  \\
   PLBART   &Encoder-Decoder &140M  &768  \\
   CodeT5   &Encoder-Decoder &220M  &768  \\
   CodeGPT  &Decoder         &124M  &768  \\
   CodeGen  &Decoder         &350M  &1024 \\

  \bottomrule

\end{tabular}
\label{tab:ptms}
\end{table*}

\subsection{Evaluation Tasks}
We choose four currently prominent SE classification tasks: code vulnerability detection (CVD), code clone detection (CCD), just-in-time software defect prediction (JIT), and function-docstring mismatch detection (FDMD). The first two tasks, code vulnerability detection and code clone detection, exclusively pertain to the source code. The latter two tasks, just-in-time software defect prediction and function-docstring mismatch detection involve both source code data and textual information like natural language annotations. These tasks are outlined briefly below.

\textbf{Code vulnerability detection} is a method used to check and discover security vulnerabilities in software systems \cite{lin-cvd, chakraborty-cvd}. Its purpose is to identify potential vulnerabilities in software code blocks that could be exploited by attackers. For instance, the CWE119 vulnerability type could allow for the execution of arbitrary code and access to sensitive information. This task has been a long-standing research focus in the SE field. In work related to vulnerability detection based on deep learning, the input typically consists of a code snippet, and the output is a label indicating the presence or absence of a vulnerability \cite{russell-cvd, li-cvd-vuldeepecker}.

\textbf{Code clone detection} involves measuring the semantic or structural similarity between two code snippets\cite{lei-ccd, tao-ccd}. The code clone technique can enhance efficiency, but it may also inadvertently introduce external vulnerabilities. Hence, code clone detection is highly essential and can have a substantial impact on practice. This task also draws attention from researchers in the SE community. In prior related studies, the input typically consists of two code snippets, and the output is a label indicating whether they are similar \cite{zhang-ccd, zakeri-ccd}.

\textbf{Just-in-time software defect prediction} aims to forecast whether a developer's commits, made during software development, will potentially introduce defects in the future\cite{zhao-jit, song-hit}. This commit-level defect prediction provides a valuable tool for testers to prioritize their limited software quality assurance resources towards the highest-risk commits. As a highly time-effective testing aid, just-in-time software defect prediction has garnered significant attention from numerous researchers. In related works on just-in-time defect prediction based on deep learning, the input typically comprises code snippets changed before and after a commit and the comments associated with that commit\cite{zeng-jit, hoang-jit, zhou-codebert4jit}.

\textbf{Function-docstring mismatch detection} is employed to assess whether the function and its associated docstring match. In the engineering practice of software development, developers are encouraged to provide descriptive natural language documentation that elucidates the purpose and usage of functions. This practice establishes a parallel corpus between code snippets and natural language sentences  \cite{husain-codesearchnet}. Many researchers evaluate various SE application tasks on such corpora, including machine translation and code search. Aditya et al. \cite{kanade-cubert} curated this data and devised a sentence pair classification task, where a function and its corresponding docstring are treated as distinct sentences. The positive instance represents the correct function-docstring pairing, while the negative instance is a function-docstring pair whose docstring is replaced by that of another function randomly selected from the dataset. Clearly, the input for this task consists of code snippets along with their corresponding natural language descriptions.

We initiate our investigation of RQ1 by examining the first three tasks introduced earlier: code vulnerability detection, code clone detection, and just-in-time software defect prediction. Although the input of the JIT task is code snippets and commit messages, the JIT task can also be completed relying only on code data, while the FDMD task cannot be performed only on code data. Consequently, in our exploration of RQ1, we concentrate on tasks where the input comprises solely code snippets, and when JIT tasks are used to study RQ1 we only input code snippets. Moving on to RQ2, we delve into two tasks: just-in-time software defect prediction and function-docstring mismatch detection, both of which require the simultaneous input of code snippets and their corresponding text information.

\begin{table}
\centering
  \caption{Code Vulnerability Detection Dataset Statistics}
  \begin{tabular}{cccc}
    \toprule
Project   & Examples & Vulnerable ratio & Language  \\
    \midrule
Devign    & 22361            & 45.02\%                  & C/C++   \\
CWE119    & 39753            & 26.26\%                  & C/C++    \\
CWE399    & 21885            & 33.29\%                  & C/C++    \\
  \bottomrule
\end{tabular}
\label{tab:cvd-datasets}
\end{table}

\begin{table}
\centering
  \caption{Code Clone Detection Dataset Statistics}
  \begin{tabular}{cccc}
    \toprule
Project   & Examples & Clone type & Language  \\
\midrule
BigCloneBench    & 1731860   & Type-1,2,3,4   & Java   \\
  \bottomrule
\end{tabular}
\label{tab:ccd-datasets}
\end{table}

\begin{table}
\centering
  \caption{Just-in-time Software Defect Prediction Dataset Statistics}
  \begin{tabular}{cccc}
    \toprule
Project   & Changes & Defect ratio & Language  \\
    \midrule
qt        & 95978            & 15.16\%                  & C++        \\
openstack & 66065            & 31.68\%                  & C++    \\
platform  & 39365            & 37.74\%                  & Java     \\
gerrit    & 34610            & 8.64\%                   & Java     \\
go        & 61224            & 36.75\%                  & Golang  \\
  \bottomrule
\end{tabular}
\label{tab:jit-datatset}
\end{table}

\begin{table}
\centering
  \caption{Function Docstring Mismatch Detection Dataset Statistics}
  \label{tab:fdmd-datasets}
  \begin{tabular}{cccc}
    \toprule
Project   & Examples & Mismatch ratio & Language  \\
\midrule
ETH-Py150    & 260892   & 50.00\%   & Python   \\
  \bottomrule
\end{tabular}
\label{tab:cc}
\end{table}

\subsection{Experimental Datasets}
We gathered the dataset necessary for this paper from the open-source community. Thanks to the spirit of open sharing, we were able to readily acquire the datasets for the four classification tasks chosen for this paper.

For the CVD task, we utilized three datasets: Devign, CWE119, and CWE399. Devign \cite{zhou-devign} consists entirely of real-world vulnerability codes, meticulously compiled by Zhou et al. CWE119 and CWE399, collected by Li et al. \cite{li-cvd-vuldeepecker}, incorporate both real-world and synthetic examples. In the case of the CCD task, we conducted experiments on the widely recognized BigCloneBench dataset \cite{svajlenko-bigclonebench}. This dataset contains four different types of Java code clones from Type-1 to Type-4 and is employed to ascertain whether two given Java code snippets share similar semantics. For the JIT task, we employed five datasets: openstack, qt, platform, gerrit, and go. These encompass three different programming languages: C++, Java, and GoLang. The dataset, compiled by Zeng et al. \cite{zeng-jit}, primarily comprises records from the software development process. It includes details of code changes, commit messages, and a label indicating whether the commit contains defects. In the case of the FDMD task, we relied on the dataset published by Aditya et al. \cite{kanade-cubert}, referred to as the ETH-Py150 dataset. This dataset predominantly features combinations of Python functions along with their corresponding correct or incorrect documentation strings. It serves as training material for a classifier aimed at distinguishing between correctly and incorrectly matched pairs.

In summary, Tables \ref{tab:cvd-datasets}-\ref{tab:fdmd-datasets} provide concise overviews of the datasets under scrutiny. It's worth noting that these chosen datasets display diversity from multiple perspectives, including programming language and dataset size. This diversity serves to guard against unstable conclusions and ensures that the results we derive can be generalized across a broader spectrum of tasks and datasets.

\subsection{Performance Metrics}
Given that this paper centers around SE-related classification tasks, we employ commonly used classification metrics from the field of machine learning to assess task performance. While the tasks mentioned earlier have been evaluated by previous researchers using only one evaluation metric, such as Accuracy often being used in vulnerability detection \cite{lu-codexglue-codegpt, wang-codet5}, relying on only one single metric can potentially lead to biased conclusions \cite{agrawal-better}. Therefore, this paper opts for three evaluation metrics: Accuracy, F1-Score, and MCC.

Accuracy, the simplest and most intuitive metric, gauges the proportion of correctly classified instances to the total number of instances. Precision and Recall correspond to the model's ability to detect accurately and completely, respectively, but these two indicators are contradictory to each other. F1-Score, calculated as the harmonic mean of Precision and Recall, strikes a balance between these two factors, offering a comprehensive assessment of the model's performance. MCC, on the other hand, takes into account all four possible outcomes in classification: TP, TN, FN, and FP, presenting a relatively balanced metric. Formulas 1-5 are used to calculate these three metrics.

\begin{equation}
\text{Accuracy} = \frac{TP+TN} {TP+TN+FP+FN}
\end{equation}

\begin{equation}
\text{Precision} = \frac{TP}{TP+FP}
\end{equation}

\begin{equation}
\text{Recall} = \frac{TP}{TP+FN}
\end{equation}

\begin{equation}
\text{F1-Score} = \frac{2 \times Precision \times Recall}{Precision + Recall}
\end{equation}

\begin{equation}
\text{MCC} = \frac{TP \times TN - FP \times FN}{(TP + FP)(TP + FN)(TN + FP)(TN + FN)}
\end{equation}

Among them, TP and TN represent correct predictions, signifying positive samples being classified as positive and negative samples as negative, respectively. Conversely, FP and FN denote prediction errors, with FP indicating negative samples being classified as positive, and FN indicating positive samples being classified as negative.

\section{Experimental Methods and Results}
\label{sec: results}

In this section, we first introduce the detailed methodological steps to explore the two research questions of this paper. We then present quantitative results on code embeddings derived from previously identified PTMs on several downstream classification tasks. Finally, based on the analysis of the experimental results, we address the two research questions that are the focus of this paper.

\subsection{RQ1: Can the embedding obtained through a particular token sufficiently aggregate the semantic information of the entire code snippets?}

\begin{figure*}[h]
  \centering
  \includegraphics[width=\linewidth]{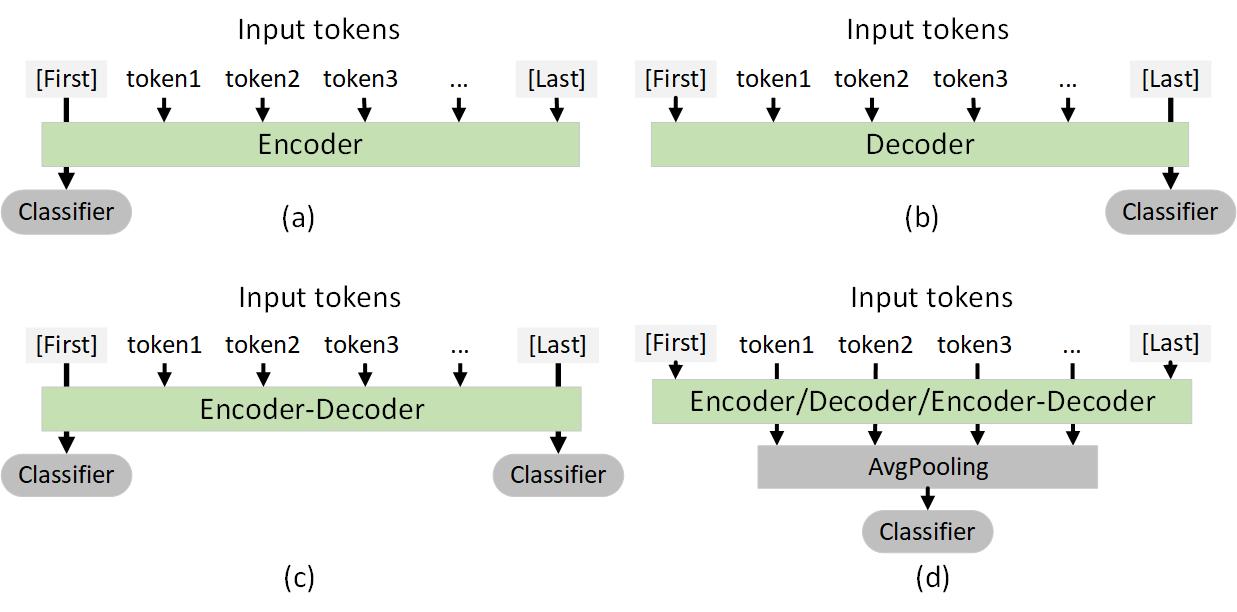}
  \caption{Different ways to obtain code embeddings of PTMs with three architectures when performing classification tasks.}
\label{fig:rq1-approach}
\end{figure*}

\textbf{Approach} To address this question and ensure the generalizability of our conclusions, we examined five code PTMs encompassing all three distinct architectures, as outlined in Table \ref{tab:ptms}. These architectures include the encoder-only architecture, decoder-only architecture, and encoder-decoder architecture.

We first utilize the method commonly employed by researchers \cite{liu-special, niu-special, chai-special, zeng-special}, as well as used when the PTM is released, to obtain the code embeddings \cite{feng-codebert, wang-codet5, lu-codexglue-codegpt}. This involves acquiring code embeddings through a specific token, as shown in subfigures (a), (b) and (c) in Figure \ref{fig:rq1-approach}. For CodeBERT, we extract the embedding of the first token (known as [CLS]) from the final hidden layer of the model to serve as the aggregated semantic representation of the input code. Similarly, for CodeGen, we utilize the embedding of the last token from the final hidden layer as the aggregated semantic representation of the input code. In the case of CodeT5, due to its encoder-decoder architecture, both the embeddings of the first and last tokens from the final hidden layer of the model can aggregate the semantic representation of the input code. We then exclude these specific tokens, as illustrated in subfigure (d) in Figure \ref{fig:rq1-approach}, and adopt an alternate approach, which involves aggregating the vector representations of each input code token. We derive the semantic representation of the input code by employing a straightforward method of average-pooling the embeddings of all code tokens.

Based on the obtained aggregated semantic embeddings of the input code snippets, we construct and train a simple fully connected layer for classification. We compare the performance of the classifier to determine whether the method of using special tokens can effectively aggregate the semantic information of the entire code snippet. To mitigate performance biases resulting from experimental randomness, we conducted the experiment 50 times and then averaged the performance metrics. To ensure the significance of experimental performance, we subjected the performance value distribution to statistical testing using the Wilcoxon Signed Rank Test method \cite{woolson-wilcoxon}.

\begin{table*}[]
\fontsize{5.1}{6}\selectfont  
\caption{Evaluation results on the test set of three downstream tasks, where F, L and A respectively represent the performance of the classifier built based on the code embedding obtained by the first special token, the last special token and the average-pooling of all code tokens. The bold value indicates the optimal performance value under the same PTM.}
\setlength{\tabcolsep}{0.50 mm}{
\begin{tabular}{ccc|cc|ccc|ccc|cc|cc}

\toprule
\multirow{2}{*}{Tasks}&\multirow{2}{*}{Datasets} &\multirow{2}{*}{Metrics}& \multicolumn{2}{c}{CodeBERT}&                \multicolumn{3}{c}{CodeT5} & \multicolumn{3}{c}{PLBART} & \multicolumn{2}{c}{CodeGPT}& \multicolumn{2}{c}{CodeGen}\\
 &&                                       & F& A&F& L& A& F& L&A& L& A& L& A\\
 \midrule
 \multirow{15}{*}{JIT}& \multirow{3}{*}{go}
 & ACC & \textbf{0.619}& 0.613& 0.601& 0.600& \textbf{0.616}& 0.610& 0.602& \textbf{0.619}& 0.586& \textbf{0.612}& 0.604&\textbf{0.618}\\
& & F1  & 0.582& \textbf{0.583}& 0.542& \textbf{0.583}& 0.573& 0.566& 0.558& \textbf{0.581}& 0.561& \textbf{0.581}& 0.561&\textbf{0.574}\\
& & MCC  & \textbf{0.259}& 0.255& 0.212& 0.246& \textbf{0.246}& 0.233& 0.217& \textbf{0.257}& 0.206& \textbf{0.250}& 0.230&\textbf{0.249}\\
& \multirow{3}{*}{platform}
& ACC & 0.654& \textbf{0.665}& 0.612& 0.629&\textbf{0.663}& 0.659& 0.647& \textbf{0.669}& 0.617& \textbf{0.647}& 0.611&\textbf{0.661}\\
& & F1 & 0.587& \textbf{0.600}& 0.560& 0.586& \textbf{0.592}& 0.589& 0.592& \textbf{0.606}& 0.582& \textbf{0.598}& 0.562&\textbf{0.594}\\
& & MCC  & 0.347& \textbf{0.370}& 0.306& 0.342& \textbf{0.355}& 0.350& 0.352&\textbf{0.379}& 0.332& \textbf{0.364}& 0.312&\textbf{0.359}\\
& \multirow{3}{*}{gerrit}
& ACC  & \textbf{0.813}& 0.787& 0.760& 0.779& \textbf{0.812}& 0.806& 0.813& \textbf{0.824}& 0.743& \textbf{0.806}& 0.770&\textbf{0.825}\\
& & F1 & \textbf{0.167 }& 0.164& 0.157& 0.149& \textbf{0.178}& 0.165& 0.175& \textbf{0.187}& 0.150& \textbf{0.193}& 0.149&\textbf{0.200}\\
& & MCC & \textbf{0.097}& 0.092& 0.087 & 0.072 & \textbf{0.111} & 0.094 & 0.107 & \textbf{0.122} & 0.074 & \textbf{0.129} & 0.074 &\textbf{0.137} \\
& \multirow{3}{*}{openstack}
& ACC & \textbf{0.585} & 0.574& 0.572& 0.567 & \textbf{0.599} & 0.583 & 0.616 &\textbf{0.628} & 0.553 & \textbf{0.603} & 0.579 &\textbf{0.621} \\
& & F1 & 0.388 & \textbf{0.392} & 0.377 & 0.383 &\textbf{0.390} & 0.379 & 0.368 & \textbf{0.395} & 0.368 & \textbf{0.401} & 0.379 &\textbf{0.398} \\
& & MCC  & 0.195 & \textbf{0.203} & 0.182 & 0.186 & \textbf{0.198} & 0.180 & 0.166 & \textbf{0.208} & 0.159 & \textbf{0.217} & 0.186 &\textbf{0.212} \\
& \multirow{3}{*}{qt}
& ACC & \textbf{0.638} & 0.623 & 0.582 & 0.606 & \textbf{0.672} & 0.637 & 0.675 & \textbf{0.672} & 0.607 & \textbf{0.644} & 0.584 &\textbf{0.683} \\
& & F1 & \textbf{0.337} & 0.335 & 0.330 & \textbf{0.342} & 0.340 & 0.312 & 0.321 & \textbf{0.339} & 0.325 & \textbf{0.333} & 0.336 &\textbf{0.339} \\
& & MCC & \textbf{0.196} & 0.195 & 0.194 & \textbf{0.210} & 0.200 & 0.156 & 0.1727 & \textbf{0.197} & 0.178 & \textbf{0.190} & \textbf{0.205} &0.198\\

\multirow{9}{*}{CVD}&\multirow{3}{*}{Devign} 
& ACC & \textbf{0.568} & 0.561 &0.546 & 0.557 & \textbf{0.593}  & 0.571 & 0.549 & \textbf{0.592} & 0.531 &\textbf{0.580} & 0.557 & \textbf{0.599} \\
&& F1  & \textbf{0.551} & 0.542  &0.510 & 0.540  & \textbf{0.573}  & 0.551 & 0.523 & \textbf{0.579} & 0.504 & \textbf{0.569} & 0.506 & \textbf{0.587} \\
&& MCC & \textbf{0.139} & 0.125 &0.093 & 0.118 & \textbf{0.187}  & 0.142 & 0.096 & \textbf{0.187} & 0.061 & \textbf{0.164} & 0.111  & \textbf{0.202} \\
&\multirow{3}{*}{CWE119} 
& ACC  & 0.758 & \textbf{0.766} &0.561 & 0.790 & \textbf{0.879}  & 0.800 & 0.747 & \textbf{0.857} & 0.640 & \textbf{0.769} & 0.647 & \textbf{0.865} \\
&& F1 & 0.720 & \textbf{0.736} &0.600 & 0.740  & \textbf{0.826}  & 0.735 & 0.678 & \textbf{0.817} & 0.605 & \textbf{0.705} & 0.660 & \textbf{0.820} \\
&& MCC  & 0.542& \textbf{0.569} &0.298 & 0.578 & \textbf{0.737}  & 0.579 & 0.478 & \textbf{0.707} & 0.325 & \textbf{0.523} & 0.437  & \textbf{0.716} \\
&\multirow{3}{*}{CWE399} 
& ACC  & 0.704   & \textbf{0.742} &0.694 & \textbf{0.751} & 0.735  & 0.756 & 0.754 & \textbf{0.782} & \textbf{0.745} & 0.734 & 0.708 & \textbf{0.740} \\
&& F1  & \textbf{0.416} & 0.408 &0.272 & 0.393   & \textbf{0.479}  & 0.447 & 0.405 & \textbf{0.525} & 0.374 & \textbf{0.464} & 0.356 & \textbf{0.483} \\
&& MCC & 0.236  & \textbf{0.249} & 0.088 & 0.238  & \textbf{0.319}  & 0.295& 0.252& \textbf{0.389}& 0.216& \textbf{0.300}& 0.192 & \textbf{0.325} \\

 \multirow{3}{*}{CCD}&\multirow{3}{*}{BigCloneBench} 
& ACC & 0.687 & \textbf{0.722} & 0.635 & 0.665 & \textbf{0.726}  & \textbf{0.735} & 0.675&0.727 & 0.640 & \textbf{0.911}& 0.582 & \textbf{0.930} \\
&& F1 & 0.403 & \textbf{0.427} & 0.376 & 0.395 & \textbf{0.449}  & 0.427 & 0.379 & \textbf{0.455} & 0.355& \textbf{0.730}& 0.346 & \textbf{0.778} \\
&& MCC & 0.310 & \textbf{0.343} & 0.286& 0.308 & \textbf{0.378}  & 0.337 & 0.277 & \textbf{0.388} & 0.244& \textbf{0.692}& 0.248 & \textbf{0.747} \\
\bottomrule
\end{tabular}
}
 \label{tab:rq1-performance-results}
\end{table*}

\begin{table*}[]
\tiny
\caption{Statistical test results of the performance of classifiers built based on code embeddings obtained in different ways, where F\&A represents the performance comparison of the first special token and the average-pooling of all code tokens. L\&A is similar. \% represents the performance improvement percentage of the average-pooling of all code tokens relative to other methods, and p represents the p-value of the significance test. \# represents $p \geq 0.05$, * represents $p < 0.05$, ** represents $p < 0.01$, *** represents $p < 0.001$. Bold values represent the maximum performance improvement ratio for each row.}
\setlength{\tabcolsep}{0.50 mm}{
\begin{tabular}{ccc|cc|cccc|cccc|cc|cc}
\toprule
 \multirow{2}{*}{Tasks}
&\multirow{2}{*}{Datasets}&\multirow{2}{*}{Metrics}&  \multicolumn{2}{c}{CodeBERT}& \multicolumn{4}{c}{CodeT5}&  \multicolumn{4}{c}{PLBART}&  \multicolumn{2}{c}{CodeGPT}&  \multicolumn{2}{c}{CodeGen}\\
&& &  \multicolumn{2}{c}{F\&A}& \multicolumn{2}{c}{F\&A}& \multicolumn{2}{c}{L\&A}& \multicolumn{2}{c}{F\&A}& \multicolumn{2}{c}{L\&A}& \multicolumn{2}{c}{L\&A}& \multicolumn{2}{c}{L\&A}\\
 & & & \%& p& \%& p& \%& p& \%& p& \%& p& \%& p& \%&p\\
\midrule
\multirow{15}{*}{JIT}
&\multirow{3}{*}{go}
& ACC  &  -0.90&\#& 2.61&***  &  2.73&***  &  1.59&***&  2.89&***&  \textbf{4.50}&***&  2.26&***  \\                   
&& F1 &  0.19&\#  & \textbf{5.68}&***  &  -1.75&*** &  2.58&***&  4.12&***&  3.53&***&  2.35&*\\                       
&& MCC &  -1.50&\#& 15.99&***  &  0.24&\#&  10.30&***&  18.32&***&  \textbf{21.69}&***&  8.07&***     \\
&\multirow{3}{*}{platform}
& ACC  &  1.62&**& \textbf{8.38}&***  &  5.40&***  &  1.53&***&  3.52&***&  4.94&***&  8.17&***  \\                    
&& F1 &  2.23&***& 5.69&***  &  1.01&*** &  2.83&***&  2.35&***&  2.78&***&  \textbf{5.84}&***     \\                  
&& MCC &  6.59&***  & \textbf{15.97}&***  &  3.85&*** &  8.28&***&  7.48&***&  9.50&***&  14.80&**\\
&\multirow{3}{*}{gerrit}
& ACC  &  -3.12&***  & 6.89&***  &  4.30&***  &  2.26&***&  1.35&***&  \textbf{8.52}&***&  7.16&***  \\                
&& F1 &  -2.09&\#  & 13.37&***  &  19.99&*** &  13.33&***&  7.00&***&  28.07&***&  \textbf{34.00}&***     \\           
&& MCC &  -4.43&\#& 27.52&***  &  53.17&*** &  28.96&***&  14.02&***&  72.86&***&  \textbf{84.45}&***     \\
&\multirow{3}{*}{openstack}
& ACC  &  -1.91&*& 4.68&*&  5.58&***  &  7.66&***&  2.00&**&  \textbf{9.07}&***&  7.32&***  \\                         
&& F1 &  1.16&**& 3.47&***  &  1.93&*** &  4.13&***&  7.24&***&  \textbf{9.07}&***&  4.98&***     \\                   
&& MCC &  4.36&***  & 8.81&***  &  6.60&*** &  15.43&***&  25.00&***&  \textbf{36.52}&***&  13.96&***     \\
&\multirow{3}{*}{qt}
& ACC  &  -2.29&*& 15.44&***  &  10.76&***  &  5.46&***&  -0.49&\#&  6.05&***&  \textbf{16.93}&***  \\                
&& F1 &  -0.44&\#& 3.12&***  &  -0.61&*&  \textbf{8.58}&***&  5.38&***&  2.58&***&  1.13&***     \\                    
&& MCC &  -0.51&***& 2.88&\#&  -4.76&*** &  \textbf{25.69}&***&  14.19&***&  6.44&***&  -3.50&***     \\
\multirow{9}{*}{CVD}
&\multirow{3}{*}{Devign}
& ACC  &  -1.25&***  & 8.68&***  &  6.44&***  &  3.75&***&  7.86&***&  \textbf{9.23}&***&  7.52&***  \\                
&& F1 &  -1.70&\#  & 12.16&***  &  6.09&*** &  5.06&***&  10.67&***&  13.03&***&  \textbf{15.84}&***     \\            
&& MCC &  -10.04&***  & 100.97&***  &  58.33&*** &  31.51&***&  93.60&***&  \textbf{168.57}&***&  81.00&***     \\
&\multirow{3}{*}{CWE119}
& ACC  &  1.09&\#  & \textbf{56.75}&***  &  11.23&*** &  7.13&***&  14.71&***&  20.02&***&  33.83&*** \\              
&& F1 &  2.17&\#   & \textbf{37.65}&*** &  11.63&*** &  11.22&***&  20.60&***&  16.52&***&  24.16&***  \\            
&& MCC&  4.86&\#  & \textbf{147.08}&*** &  27.56&*** &  22.16&***&  48.02&***&  61.01&***&  63.94&***  \\
&\multirow{3}{*}{CWE399}
& ACC &  5.45&***  & \textbf{5.89}&***  &  -2.05&*** &  3.48&***&  3.81&***&  -1.46&**&  4.59&***     \\          
&& F1 &  -1.92&\#  & \textbf{75.84}&*** &  22.06&***  &  17.36&***&  29.42&***&  24.09&***&  35.55&***     \\  
&& MCC &  5.24&\# & \textbf{262.36}&*** &  33.78&*** &  31.72&***&  54.14&***&  39.03&***& 69.32&***  \\
\multirow{3}{*}{CCD}&\multirow{3}{*}{BigCloneBench}
& ACC &  5.15&** & 14.26&*** &  9.15&***  &  -1.10&**&  7.64&***&  42.26&***&  \textbf{59.75}&*** \\
&& F1 &  5.89&*** & 19.53&*** &  13.75&***  &  6.40&***&  19.94&***&  105.55&***&  \textbf{125.09}&*** \\
&& MCC &  9.54&*** & 32.48&*** &  22.74&***  &  14.92&***&  39.94&***&  183.63&***& \textbf{200.36}&***\\
\bottomrule
\end{tabular}
}
 \label{tab:rq1-significance-test}
\end{table*}

\begin{figure}[h]
  \centering
  \includegraphics[width=\linewidth]{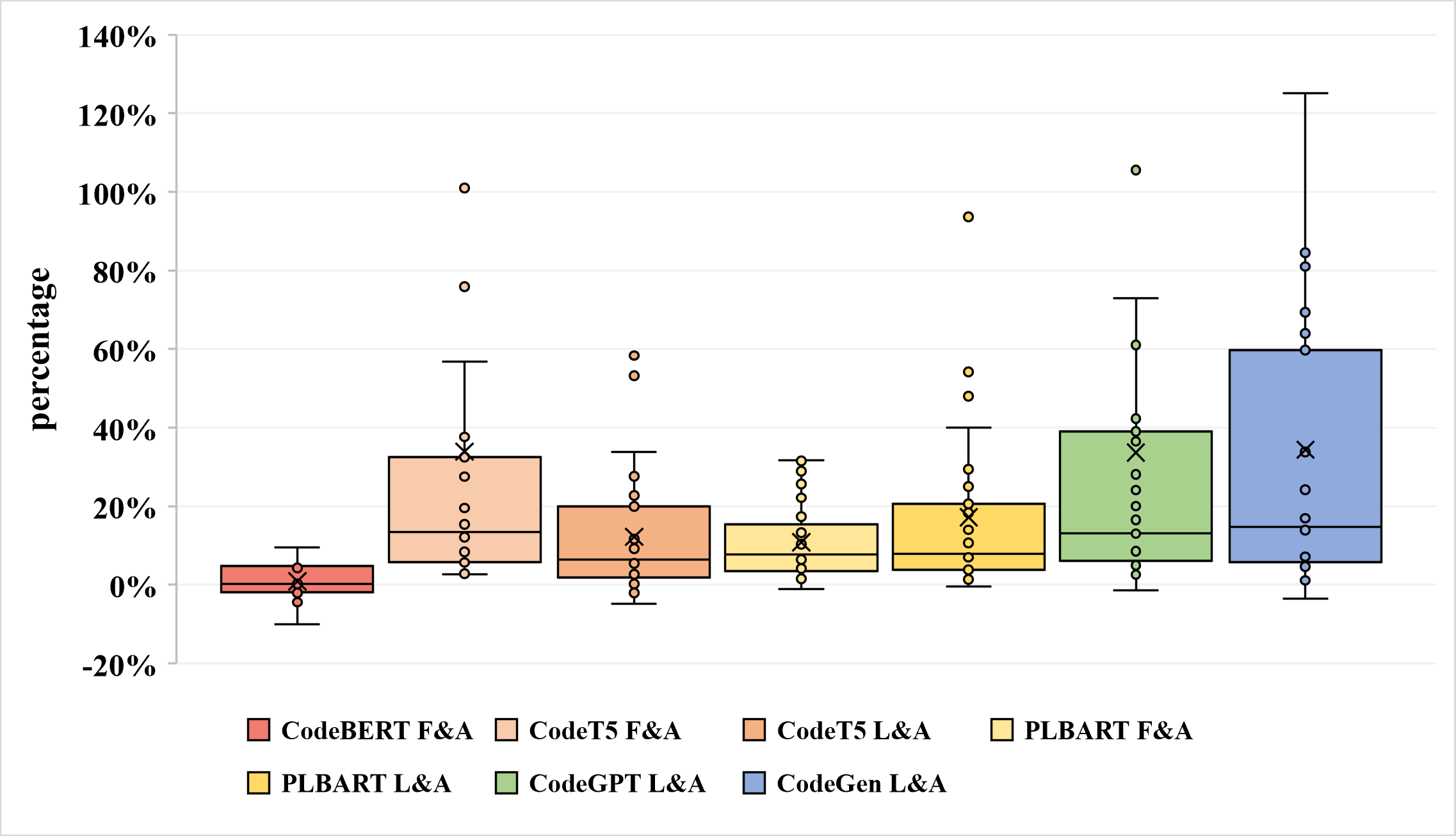}
  \caption{Illustration of the performance improvement percentage of the classifier constructed based on the code embeddings obtained by the average-pooling of all code tokens compared to the special token.}
\label{fig:rq1-performance-percentage}
\end{figure}

\textbf{Results} Table \ref{tab:rq1-performance-results} displays the performance of three indicators on three downstream SE classification tasks for five code PTMs with three different architectures using code embeddings obtained in different ways. The bolded values represent the best performance values obtained using different methods to obtain code embedding under the same code PTM. Table \ref{tab:rq1-significance-test} shows the p-value of the significance test between the performance of the same code PTM using code embedding obtained in different ways, as well as the percentage improvement in three indicators using the average-pooling method to get the embedding. Figure \ref{fig:rq1-performance-percentage} visualizes the percentage improvements from Table \ref{tab:rq1-significance-test} for a clearer comparative observation.

\textbf{Regardless of the architecture of the code PTMs, embeddings obtained through a specific token do not sufficiently aggregate the semantic information of the entire code snippet. Instead, focusing on the vector representation of each code token, such as through a simple average pooling of all code tokens, leads to more enriched code embeddings with richer semantic information.} In Table \ref{tab:rq1-performance-results}, out of a total of 27 records across the three metrics on nine datasets for three classification tasks, CodeBERT achieved the best performance 14 times when obtaining code embeddings through average-pooling of all code tokens, while aggregate embeddings using the first special token yielded the best performance 13 times. Considering the significance test results from Table \ref{tab:rq1-significance-test}, it can be observed that average-pooling of all code tokens led to significantly better performance in 9 cases, whereas the special token approach only resulted in significantly better performance 4 times. In the remaining 14 experiments, the two methods of obtaining code embeddings can be considered equivalent. For PLBART, CodeGPT, and CodeGen, embeddings obtained through average-pooling demonstrated significantly better performance in 26 cases, while special token embeddings only outperformed once. In the case of PLBART, embeddings obtained through the last special token consistently performed the worst. For CodeT5, embeddings obtained through the last special token only outperformed four times, while those obtained through the first special token consistently performed the worst. On the other hand, embeddings obtained through average-pooling showed significantly better performance in 22 cases. This indicates that code PTMs differ from models used in the NLP field. The common NLP approach of obtaining embeddings through special tokens does not effectively aggregate the semantic information of code snippets. Instead, employing a straightforward method like average-pooling across all code tokens results in higher-quality code embeddings, consequently enhancing the performance of SE classification tasks.

\textbf{The quality of code embeddings obtained through different methods of PTMs is influenced by the code data of downstream classification tasks.} If the PTM includes the programming language of the downstream task during its pre-training, then the quality of the semantic embeddings obtained by aggregating the vector representations of all code tokens is higher. If this language is not included, then the quality of code embeddings obtained through special tokens and those obtained by aggregating the vector representations of all code tokens is similar. Taking the CVD task and CCD task as examples, the codes of the three datasets of the CVD task are composed of C/C++ language, and the dataset of the CCD task is written in Java language. The encoder-only architecture of CodeBERT, during pre-training, includes six programming languages: Python, Java, JavaScript, PHP, Ruby, and Golang, but does not include C and C++. Therefore, we observe that CodeBERT exhibits a comprehensive and significant advantage in code clone detection tasks when classifiers are constructed based on embeddings obtained through average-pooling of all code tokens, compared to embeddings obtained through special tokens. In the code vulnerability detection task, embeddings obtained by average-pooling of all code tokens only demonstrate a significant advantage in the Accuracy of CWE399, while showing a significant disadvantage in the MCC of Devign. In the remaining code vulnerability detection datasets and metrics, code embeddings obtained through average-pooling of all code tokens do not exhibit a significant difference in quality compared to those obtained with special tokens. For code PTMs like CodeT5 and CodeGen, these models use C and C++ languages during pre-training. Consequently, the semantic embeddings acquired by averaging pooling of all code tokens contain richer information, resulting in better model performance for classifiers constructed based on these embeddings.

\textbf{Aggregating the vector representations of all code tokens to obtain semantic embeddings is more effective for PTMs of encoder-decoder and decoder-only architectures, while PTMs of decoder-only architecture benefit the most.} As shown in Table \ref{tab:rq1-significance-test} and Figure \ref{fig:rq1-performance-percentage}, for the encoder-only architecture of CodeBERT, compared to embeddings obtained through special tokens, the performance improvement of classifiers constructed based on embeddings obtained through average-pooling is within 10\% across all three metrics. In contrast, for PTMs with encoder-decoder and decoder-only architectures, classifiers constructed based on embeddings obtained through average-pooling generally exhibit higher performance improvements across all three metrics. Among all the experimental records, the experimental records with higher performance improvement ratios all appear in the PTMs of encoder-decoder architecture and the decoder-only architecture, and some performance improvement ratios even reach more than 200\%. For instance, on the CWE399 dataset in the code vulnerability detection task, the classifier constructed based on CodeT5's embeddings obtained through average-pooling shows a 262.36\% improvement in the MCC metric compared to the classifier constructed based on embeddings obtained through the first special token. Similarly, on the BigCloneBench dataset in the code clone detection task, the classifier constructed based on CodeGen's embeddings obtained through average-pooling exhibits a 200.36\% improvement in the MCC metric compared to the classifier constructed based on embeddings obtained through the last special token. However, among the 27 records, there are 16 records where CodeGPT and CodeGen of the decoder-only architecture achieved the highest performance improvement ratio, among which CodeGPT and CodeGen each achieved the highest performance improvement ratio 8 times. The remaining 11 records are the highest performance improvement ratios achieved by the encoder-decoder architecture CodeT5 and PLBART, of which CodeT5 contributed 9 times and PLBART contributed 2 times. Therefore, while all three architecture types of PTMs benefit from aggregating the vector representations of all code tokens to obtain semantic embeddings, the encoder-decoder and decoder-only architectures gain more, with decoder-only architectures benefiting the most significantly.

\textbf{By utilizing the aggregation of vector representations from all code tokens to obtain semantic embeddings, code PTMs of the decoder-only architecture can achieve embeddings that are equally rich in semantic information as those obtained from the encoder-only or encoder-decoder architecture, and in some cases, even of higher quality.} Currently, researchers tend to prefer using encoder-only architecture code PTM rather than decoder-only architecture for generating embeddings when working on SE classification tasks. This preference may stem from previous experiences where embeddings obtained using the special token method in decoder-only architecture PTMs did not encapsulate as rich semantic information as those obtained from encoder-only and encoder-decoder architectures. For instance, as shown in Table \ref{tab:rq1-performance-results}, classifiers constructed based on code embeddings derived from CodeBERT using the special token method outperformed other PTMs across most tasks and metrics. Even the larger-scale decoder-only architecture PTMs did not yield higher-quality embeddings, as seen in the case of CodeGen. However, when we adopt the strategy of average-pooling the embeddings of all code tokens, the scenario changes. Classifiers built on embeddings generated by decoder-only architecture PTMs can outperform those built on encoder-only architecture and encoder-decoder architecture. Moreover, this advantage may potentially widen with the increase in model scale. For instance, in the task of code clone detection, classifiers constructed using the average-pooling embeddings of all code tokens from CodeGPT and CodeGen outperform CodeBERT, PLBART, and CodeT5 across all metrics. This illustrates that semantic embeddings obtained through the aggregation of vector representations from all code tokens allow the decoder-only architecture PTMs to acquire code embeddings that are equally rich in semantic information as those obtained from the encoder-only or encoder-decoder architecture and even better. The fundamental reason behind this is that both decoder-only and encoder-oly architecture PTMs employ the same self-attention layer to encode word tokens, endowing them with the capability to generate semantic embeddings for text. Considering that decoder-only architecture PTMs have become the current mainstream, with increasing scales \cite{touvron-llama, dale-llm}, we recommend researchers use decoder-only architecture code PTMs for generating embeddings of code snippets.

\begin{figure*}  
	\centering  
	\vspace{-0.2cm} 
	\subfigtopskip=4pt 
	\subfigbottomskip=2pt 
	\subfigcapskip=-2pt 
	\subfigure[CodeBERT]{
		\label{fig:discussion-rq1.codebert}
		\includegraphics[width=\linewidth]{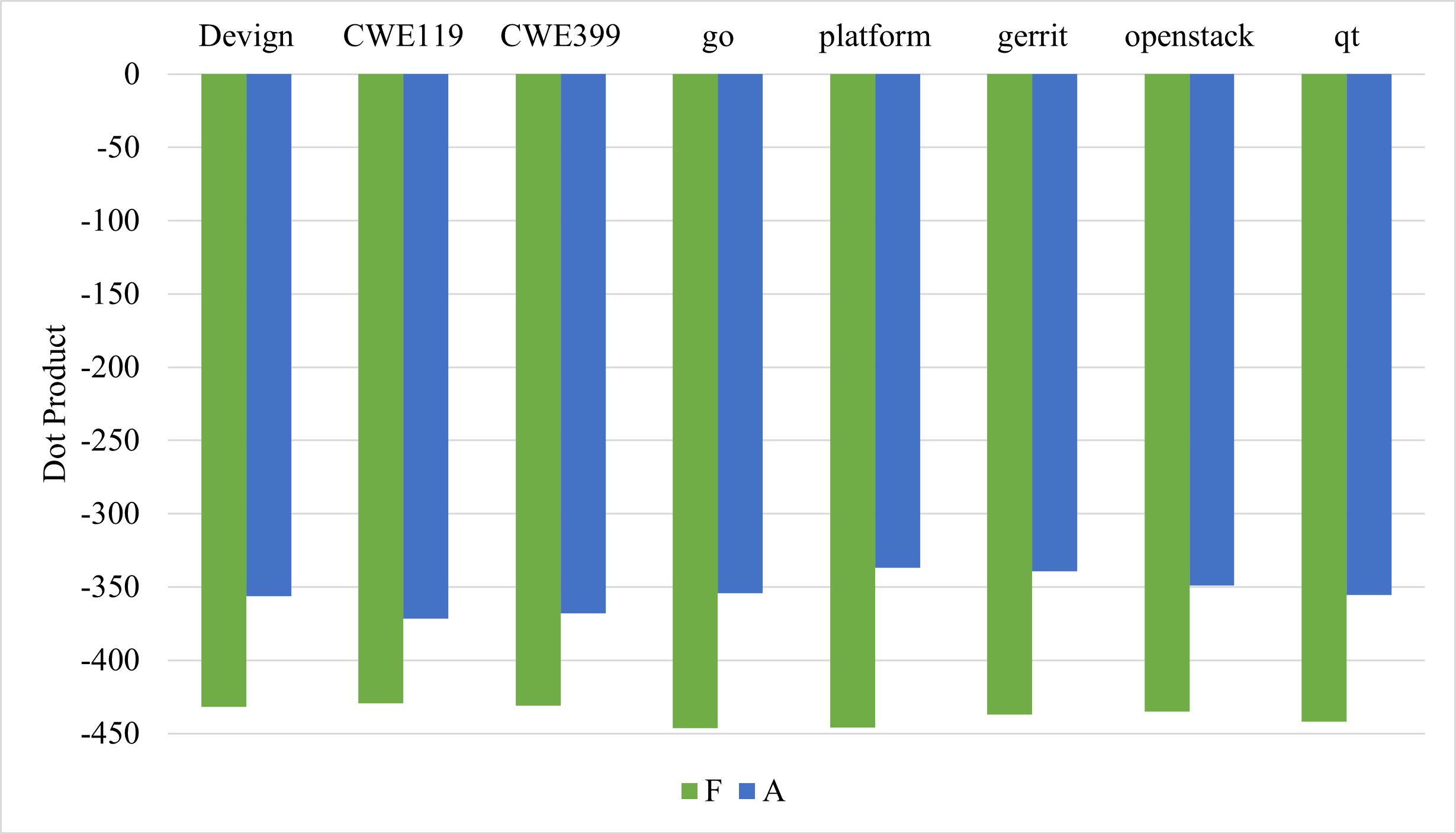}}
    \subfigure[CodeT5]{
		\label{fig:discussion-rq1.codet5}
		\includegraphics[width=\linewidth]{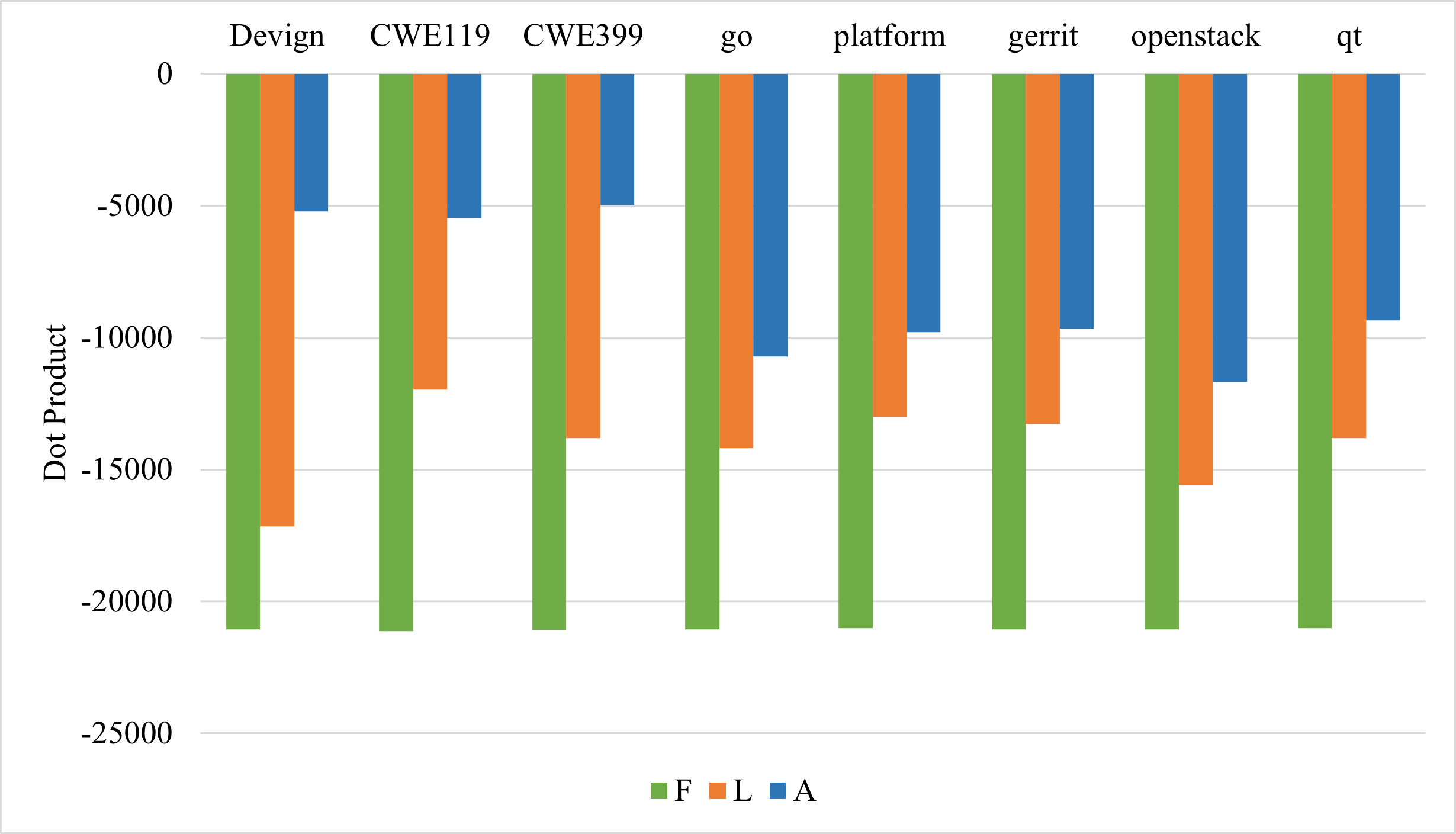}}
	\subfigure[CodeGen]{
		\label{fig:discussion-rq1.codegen}
		\includegraphics[width=\linewidth]{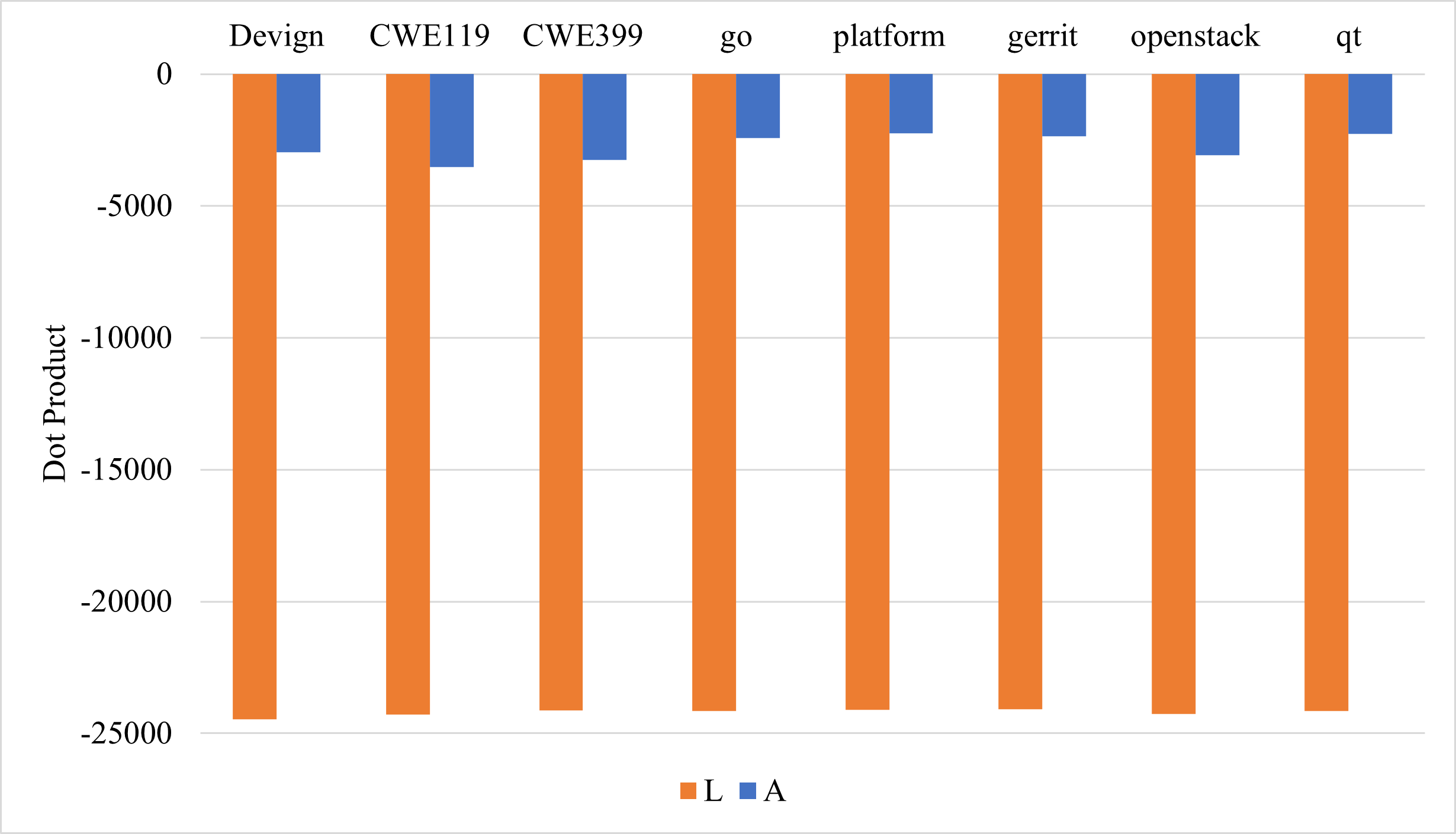}}
   \caption{Negative dot product values between code embeddings of different categories in eight projects obtained using special tokens or average-pooling all code tokens of three different architecture PTMs. Among them, subfigures (a), (b) and (c) represent encoder-only CodeBERT, encoder-decoder CodeT5 and decoder-only CodeGen respectively.}
   \label{fig:discussion-rq1}
\end{figure*}

\textbf{Discussion} By comparing the performance of classifiers constructed based on code embeddings obtained through different methods, we confirm that aggregating vector representations of all code tokens leads to embeddings with richer semantic information compared to using special tokens across all three architectures of code PTMs. One possible explanation for this phenomenon is that the method of obtaining code embeddings through special tokens fails to effectively distinguish the classification boundaries of the dataset, while embeddings obtained by aggregating vector representations of all code tokens are of higher quality and can effectively discern the boundaries between different categories. 

To investigate further, we employ the previous two different methods to obtain code embeddings for different categories of data in the test dataset and calculate the distances between code embeddings of different categories in the test set. Figure \ref{fig:discussion-rq1} displays the negative dot product values between code embeddings of different categories in the test set obtained using two different methods for three different architecture PTMs (namely CodeBERT, CodeT5, and CodeGen). This metric is used to measure the similarity between different embeddings, with smaller negative dot product values indicating closer distribution. We observe that the negative dot product values between code embeddings of different categories in the test set obtained through special tokens are smaller, implying that the embeddings obtained through special tokens for different categories exhibit similar distributions. On the other hand, the negative dot product values between code embeddings of different categories in the test set obtained through aggregating vector representations of all code tokens, i.e., average-pooling, are relatively larger. This suggests that the code embedding obtained by average-pooling of all code tokens maps different categories of data into the same high-dimensional space that is far away, while the code embedding obtained by special tokens cannot effectively distinguish samples of different categories. 

\textit{\textbf{Aggregating vector representations of all code tokens, such as simple average-pooling, results in more semantically informative code embeddings on code PTMs of all three architectures compared to embeddings aggregated by a particular token. Among them, the encoder-only architecture of PTMs benefits the least, followed by the encoder-decoder architecture, while the decoder-only architecture benefits the most.}}

\subsection{RQ2: How do the way code and text are combined affect the quality of semantic embeddings generated by code pre-trained models?}

\begin{figure*}
  \centering
  \includegraphics[width=\linewidth]{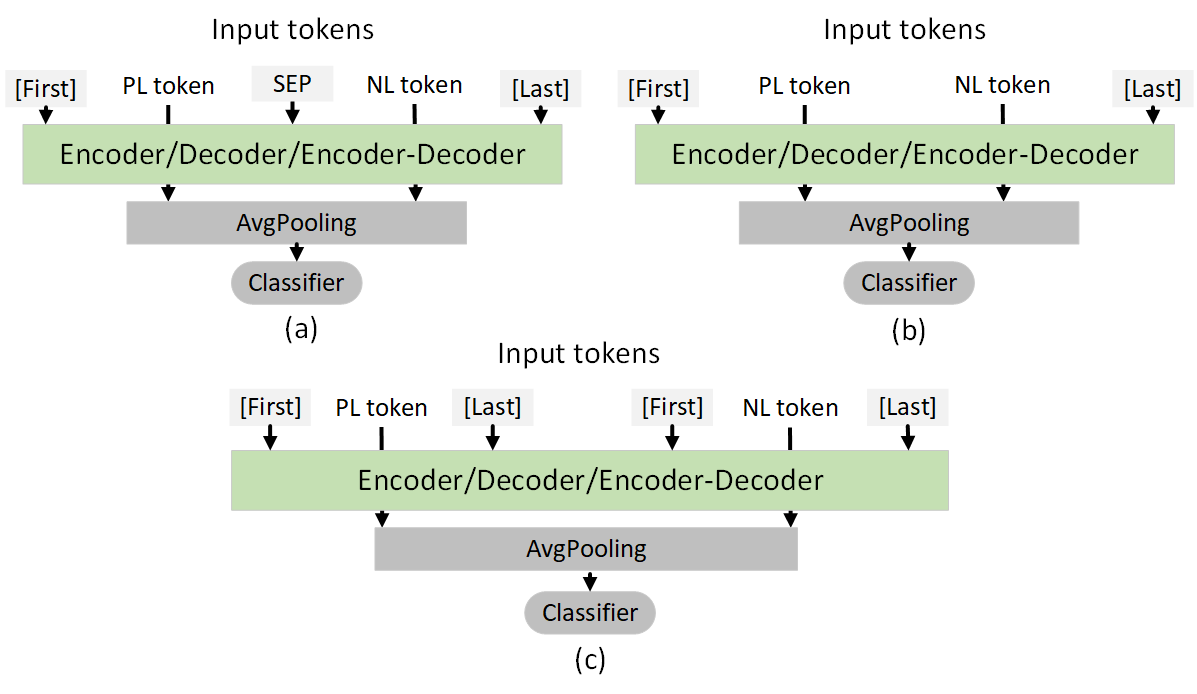}
  \caption{Three ways of combining code and text into PTMs to obtain embeddings when performing classification tasks. Among them, subfigures (a), (b) and (c) represent bimodal input, concatenated input and unimodal input respectively.}
\label{fig:rq2-approach}
\end{figure*}

\textbf{Approach} To address this question, we conduct a study on six datasets of two SE classification tasks, as shown in Tables \ref{tab:jit-datatset} and \ref{tab:fdmd-datasets}. These two classification tasks involve both code and text data as inputs. In the JIT task, the text information serves as auxiliary data, meaning that the JIT task can still be performed without this text information. On the other hand, for the FDMD task, both the text and code information are equally crucial, and both are required to complete the task. Similar to RQ1, we also investigate all five code PTMs with three different architectures, as presented in Table \ref{tab:ptms}.

The three different architecture code PTMs have variances in how they combine code data and corresponding text data during pre-training. For the encoder-only model CodeBERT and the encoder-decoder models CodeT5 and PLBART, they adopt a pairing approach during pre-training for code data with accompanying text information. This means that code and text are input together with a special token acting as a delimiter, referred to as a "bimodal input". Additionally, they also support inputting code without paired natural language text, and vice versa, known as a "unimodal input" \cite{feng-codebert, wang-codet5}. Therefore, code data with accompanying text information can be input separately as code and text, and then integrated after obtaining embeddings. For the decoder-only models CodeGPT and CodeGen, there is no special treatment for code data with accompanying text during pre-training. Instead, these data are uniformly inputted into the model and exhibit an interleaved pattern of natural language and programming language, referred to as a "concatenated input" \cite{nijkamp-codegen}. Figure \ref{fig:rq2-approach} illustrates the three different ways of combining code data and corresponding text information. We apply all three combination methods as inputs to each PTM of all architecture to obtain different embeddings.

Since the experimental results of RQ1 show that embeddings obtained by aggregating the vector representations of all code tokens contain richer semantic information, similar to RQ1, we exclude some special tokens and use a simple average-pooling method to obtain embeddings. To ensure an equal amount of input information, we control the lengths of both code and text to be the same. Based on the obtained embeddings, we construct and train a simple fully connected layer for classification. We analyze how the combination of code and text influences the quality of generated embeddings by comparing the performance of the classifiers. Similar to the process in RQ1, in order to reduce performance bias caused by experimental randomness, we conduct 50 experiments and then take the average values of the performance metrics. Additionally, to ensure the variability between experimental performances, we also use the Wilcoxon Signed Rank Test to perform a significance test on the data distribution of performance values.

\textbf{Results} Table \ref{tab:rq2-performance} illustrates the performance of classifiers constructed from embeddings obtained through various combination methods of input code and text data for five code PTMs across two downstream classification tasks, spanning three different architecture models. Bolded values indicate the best performance achieved using different combination methods for the same code PTM.
Table \ref{tab:rq2-test} showcases the p-values from significance tests comparing the performance distributions of classifiers constructed from embeddings obtained through different combination methods of input code and text for the same code PTM.

\begin{table*}
\fontsize{4.9}{6}\selectfont  
\caption{Evaluation results on the test sets of two downstream tasks, where C, B and U represent the performance of the classifier built based on the embeddings obtained from the code information and text information of the concatenated input, bimodal input and unimodal input, respectively. Bold values represent the best performance values for the same PTM.}
\setlength{\tabcolsep}{0.2 mm}{
\begin{tabular}{ccc|ccc|ccc|ccc|ccc|ccc}
\toprule
\multirow{2}{*}{Tasks}&\multirow{2}{*}{Datasets} &\multirow{2}{*}{Metrics}& \multicolumn{3}{c}{CodeBERT}&                \multicolumn{3}{c}{CodeT5} & \multicolumn{3}{c}{PLBART} & \multicolumn{3}{c}{CodeGPT}&\multicolumn{3}{c}{CodeGen}\\
 && & C&B& U&C&B& U& C&B&U& C&B& U&C&B& U\\
 \midrule
 \multirow{15}{*}{JIT}& \multirow{3}{*}{go}
 & ACC & 0.618 & 0.619 & \textbf{0.639} & 0.620 & 0.623 & \textbf{0.637} & 0.626 & 0.626 & \textbf{0.639} & 0.624 & 0.627 & \textbf{0.650} & 0.627 & 0.636 & \textbf{0.645} \\
& & F1  & 0.580 & 0.584 & \textbf{0.596} & 0.578 & 0.576 & \textbf{0.584} & 0.585 & 0.587 & \textbf{0.594} & 0.582 & 0.584 & \textbf{0.603} & 0.578 & 0.587 & \textbf{0.597} \\
& & MCC  & 0.257 & 0.262 & \textbf{0.292} & 0.255 & 0.255 & \textbf{0.278} & 0.268 & 0.269 & \textbf{0.288} & 0.262 & 0.267 & \textbf{0.307} & 0.261 & 0.279 & \textbf{0.297} \\
& \multirow{3}{*}{platform}
& ACC & 0.644 & 0.643 & \textbf{0.662} & 0.639 & 0.636 & \textbf{0.655} & 0.641 & 0.643 & \textbf{0.657} & 0.624 & 0.629 & \textbf{0.632} & 0.650 & 0.655 & \textbf{0.674} \\
& & F1 & 0.582 & 0.583 & \textbf{0.587} & 0.578 & 0.579 & \textbf{0.583} & 0.580 & \textbf{0.584} & 0.568 & 0.587 & \textbf{0.588} & 0.586 & 0.588 & \textbf{0.592} & 0.574 \\
& & MCC  & 0.338 & 0.340 &\textbf{0.350} & 0.328 & 0.329 & \textbf{0.342} & 0.331 & \textbf{0.338} & 0.320 & 0.341 & \textbf{0.345} & 0.341 & 0.347 & \textbf{0.355} & 0.337 \\
& \multirow{3}{*}{gerrit}
& ACC  & 0.795 & 0.796 & \textbf{0.809} & 0.808 & 0.805 & \textbf{0.816} & 0.812 & 0.803 & \textbf{0.822} & 0.800 & 0.805 & \textbf{0.816} & 0.826 & 0.821 & \textbf{0.838} \\
& & F1 & 0.173 & 0.172 & \textbf{0.180} & \textbf{0.201} & 0.190 & 0.192 & 0.192 & \textbf{0.196} & 0.187 & 0.199 & \textbf{0.205} & 0.197 &0.215 & \textbf{0.217} & 0.208 \\
& & MCC & 0.105 & 0.103 & \textbf{0.114} & \textbf{0.141} & 0.127 & 0.128 & 0.129 & \textbf{0.134} & 0.122 & 0.138 & \textbf{0.145} & 0.135 & 0.156 & \textbf{0.159} & 0.147 \\
& \multirow{3}{*}{openstack}
& ACC & 0.588 & 0.600 & \textbf{0.600} & 0.621 & 0.611 & \textbf{0.630} & 0.630 & 0.630 & \textbf{0.643} & 0.614 & 0.620 & \textbf{0.621} & 0.625 & 0.623 & \textbf{0.640} \\
& & F1 & 0.393 & \textbf{0.394} & 0.393 & \textbf{0.402} & 0.401 & 0.397 & 0.404 & \textbf{0.406} & 0.393 & 0.401 & \textbf{0.404} & 0.398 & 0.398 & \textbf{0.407} & 0.405 \\
& & MCC  & 0.204 & \textbf{0.205} & 0.204 & \textbf{0.218} & 0.217 & 0.213 & 0.222 & \textbf{0.224} & 0.206 & 0.216 & \textbf{0.221} & 0.212 & 0.213 & \textbf{0.226} & 0.224 \\
& \multirow{3}{*}{qt}
& ACC & 0.631 & 0.629 & \textbf{0.646} & 0.672 & 0.679 & \textbf{0.696} & 0.694 & 0.695 & \textbf{0.715} & 0.668 & 0.670 & \textbf{0.706} & 0.703 & 0.698 & \textbf{0.716} \\
& & F1 & 0.326 & \textbf{0.327} & 0.323 & 0.329 & \textbf{0.333} & 0.326 & \textbf{0.330} & 0.323 & 0.308 & \textbf{0.329} & 0.325 & 0.326 & 0.342 & \textbf{0.345} & 0.337 \\
& & MCC & 0.179 & \textbf{0.181} & 0.175 & 0.183 & \textbf{0.189} & 0.181 & \textbf{0.186} & 0.175 & 0.161 & \textbf{0.183} & 0.178 & 0.182 & 0.202 & \textbf{0.207} & 0.197 \\
\multirow{3}{*}{FDMD}
& \multirow{3}{*}{ETH-Py150}
& ACC & 0.861 & 0.859 & \textbf{0.871} & 0.836 & \textbf{0.838} & 0.725 & 0.726 & \textbf{0.868} & 0.737 & 0.673 & 0.683 & \textbf{0.697} & 0.816 & 0.828 & \textbf{0.921} \\
& & F1 & 0.860 & 0.857 & \textbf{0.870} & 0.834 & \textbf{0.837} & 0.725 & 0.724 & \textbf{0.867} & 0.732 & 0.667 & 0.680 & \textbf{0.697} & 0.814 & 0.828 & \textbf{0.920} \\
& & MCC &0.725 & 0.722 & \textbf{0.746} & 0.673 & \textbf{0.677} & 0.453 & 0.454 & \textbf{0.736} & 0.477 & 0.348 & 0.367 & \textbf{0.396} & 0.633 & 0.658 & \textbf{0.843} \\

\bottomrule
\end{tabular}
}
 \label{tab:rq2-performance}
\end{table*}

\textbf{Inputting data according to how code and text information were combined during pre-training of the code PTMs can not guarantee the acquisition of code embeddings with richer semantic information.} For the encoder-only architecture CodeBERT, bimodal input is used to input code data and corresponding text information during pre-training. However, among the 18 records shown in Table \ref{tab:rq2-performance}, we found that the bimodal input approach on the CodeBERT model only achieved the best performance four times. Yet, the significance test results in Table \ref{tab:rq2-test} indicate that these four best performance values are not significantly different from the concatenated input approach. This means that for the encoder-only architecture CodeBERT, bimodal input of code and text information in the way it is pre-trained will result in poor code embedding quality. Similarly, for the decoder-only architecture models CodeGPT and CodeGen, following their pre-training method of concatenated input of code and text information did not yield the best performance. The embeddings obtained from the CodeGen model did not achieve the best performance once, and the embeddings from the CodeGPT model only achieved the best performance twice. However, these two instances of best performance values were not significantly superior to the unimodal input approach. For the encoder-decoder architecture models CodeT5 and PLBART, utilizing the bimodal input approach as per their pre-training method to input code data and corresponding text information also did not yield a dominantly significant advantage in the quality of the obtained embeddings. Among the 18 records, the bimodal input approach obtained the best performance embeddings from the CodeT5 model only five times. While the PLBART model achieved the best performance nine times, Table \ref{tab:rq2-test} shows that four of these performance values were not significantly different from the concatenated approach. Similarly, only five instances of the best performance values were significantly superior. This implies that the proportion of classifiers constructed from embeddings obtained by inputting code information and text information in a bimodal way from these two models that achieved the best performance is less than 30\%. Therefore, regardless of the architecture of the code PTMs, inputting data according to how code and text information were combined during pre-training cannot guarantee obtaining code embeddings with richer semantic information.

\begin{table*}
\tiny
\caption{Statistical test results of the performance of classifiers built based on embeddings obtained in different ways of inputting code and text, where C\&B represents the performance comparison of the concatenated input and bimodal input. C\&U and B\&U are also similar. Among them, \# represents $p \geq 0.05$, * represents $p < 0.05$, ** represents $p < 0.01$, *** represents $p < 0.001$. }
\setlength{\tabcolsep}{0.40 mm}{
\begin{tabular}{ccc|ccc|ccc|ccc|ccc|ccc}
\toprule
 \multirow{2}{*}{Tasks}
&\multirow{2}{*}{Datasets}&\multirow{2}{*}{Metrics}& \multicolumn{3}{c}{CodeBERT}& \multicolumn{3}{c}{CodeT5}& \multicolumn{3}{c}{PLBART}& \multicolumn{3}{c}{CodeGPT}& \multicolumn{3}{c}{CodeGen}\\
&& & C\&B&C\&U&B\&U&C\&B&C\&U&B\&U&C\&B&C\&U&B\&U&C\&B&C\&U&B\&U&C\&B&C\&U&B\&U\\
\midrule
\multirow{15}{*}{JIT}
&\multirow{3}{*}{go}
& ACC  & \#&***&***& \#&***& ***& \#&***& ***& *&***&***& ***&***&***\\                                  
&& F1 & \#&***&***& \#&*& **& \#&***& ***& \#&***&***& ***&***&***\\                                    
&& MCC & *&***&***& \#&***& ***& \#&***& ***& \#&***&***& ***&***&***\\
&\multirow{3}{*}{platform}
& ACC  & \#&***&***& \#&***& ***& \#&***& ***& \#&**&\#& *&***&***\\                                  
&& F1 & \#&**&\#& \#&*& *& *&***& ***& \#&\#&\#& \#&***&***\\                                    
&& MCC & \#&***&**& \#&***& ***& *&**& ***& \#&\#&\#& *&**&***\\
&\multirow{3}{*}{gerrit}
& ACC  & \#&**&*& \#&\#& **& *&*& ***& *&***&***& \#&**&***\\                                  
&& F1 & \#&***&***& ***&***& \#& \#&*& **& **&\#&**& \#&***&***\\                                    
&& MCC & \#&***&***& ***&***& \#& \#&*& ***& *&\#&**& \#&***&***\\
&\multirow{3}{*}{openstack}
& ACC  & \#&\#&\#& \#&\#& **& \#&***& ***& \#&\#&\#& \#&*&***\\                                  
&& F1 & \#&\#&\#& \#&**& *& \#&***& ***& *&*&***& ***&***&\#\\                                    
&& MCC & \#&\#&\#& \#&*& \#& \#&***& ***& *&\#&***& ***&***&\#\\
&\multirow{3}{*}{qt}
& ACC  & \#&\#&*& \#&***& **& \#&***& ***& \#&***&***& \#&**&***\\                                  
&& F1 & \#&\#&**& **&\#& ***& ***&***& ***& **&\#&\#& **&**&***\\                                    
&& MCC & \#&\#&**& **&\#& ***& ***&***& ***& **&\#&\#& **&*&***\\
\multirow{3}{*}{FDMD}
&\multirow{3}{*}{ETH-Py150}
& ACC  & \#&***&***& ***&***& ***& ***&***& ***& ***&***&***& ***&***&***\\                                  
&& F1 & \#&***&***& ***&***& ***& ***&**& ***& ***&***&***& ***&***&***\\                                    
&& MCC & \#&***&***& ***&***& ***& ***&***& ***& ***&***&***& ***&***&***\\
\bottomrule
\end{tabular}
}
 \label{tab:rq2-test}
\end{table*}

\textbf{Regardless of the architecture of the code PTMs, employing the unimodal input approach for code and text information proves to be a competitive method for obtaining higher-quality code embeddings. The bimodal input approach comes next in effectiveness, while the concatenated input approach yields code embeddings with the least amount of semantic information.} Among the 18 records shown in Table \ref{tab:rq2-performance}, the performance of the classifier built by code embedding obtained by concatenating the input on the five PTMs of the three architectures is generally low. Specifically, on CodeBERT and CodeGen, the concatenated approach did not achieve the best performance for code embeddings once. On CodeT5 and PLBART, it only achieved the best performance twice, and on CodeGPT, it achieved the best performance twice but without significance. For CodeBERT in the encoder-only architecture, the unimodal input approach achieved significantly superior performance values 11 times, while the bimodal input approach did not achieve it even once. In the case of the encoder-decoder architecture of CodeT5 and PLBART, the unimodal input approach achieved significantly superior performance values seven times, while the bimodal input approach achieved it five times. In the decoder-only architecture of CodeGPT and CodeGen, the unimodal input approach respectively achieved significantly superior performance values eight and ten times, while the bimodal input approach achieved it four and three times. Therefore, irrespective of the architecture of the pre-trained model, employing the unimodal input approach allows for the full utilization of both code and text information, resulting in code embeddings that encapsulate richer semantic information.

\begin{figure*}  
	\centering  
	\vspace{-0.2cm} 
	\subfigtopskip=4pt 
	\subfigbottomskip=2pt 
	\subfigcapskip=-2pt 
	\subfigure[CodeBERT]{
		\label{fig:discussion-rq2.codebert}
		\includegraphics[width=\linewidth]{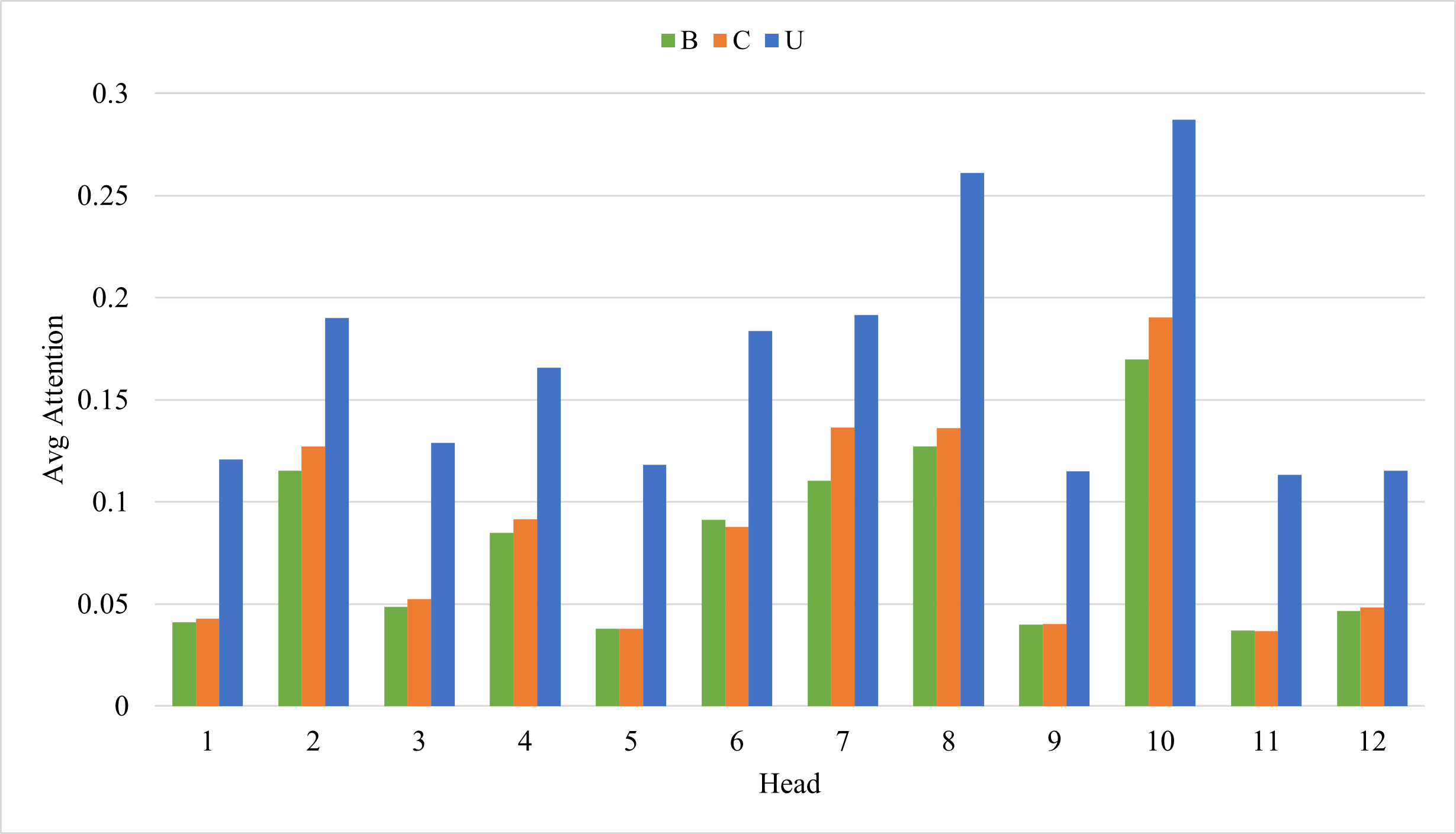}}
    \subfigure[CodeT5]{
		\label{fig:discussion-rq2.codet5}
		\includegraphics[width=\linewidth]{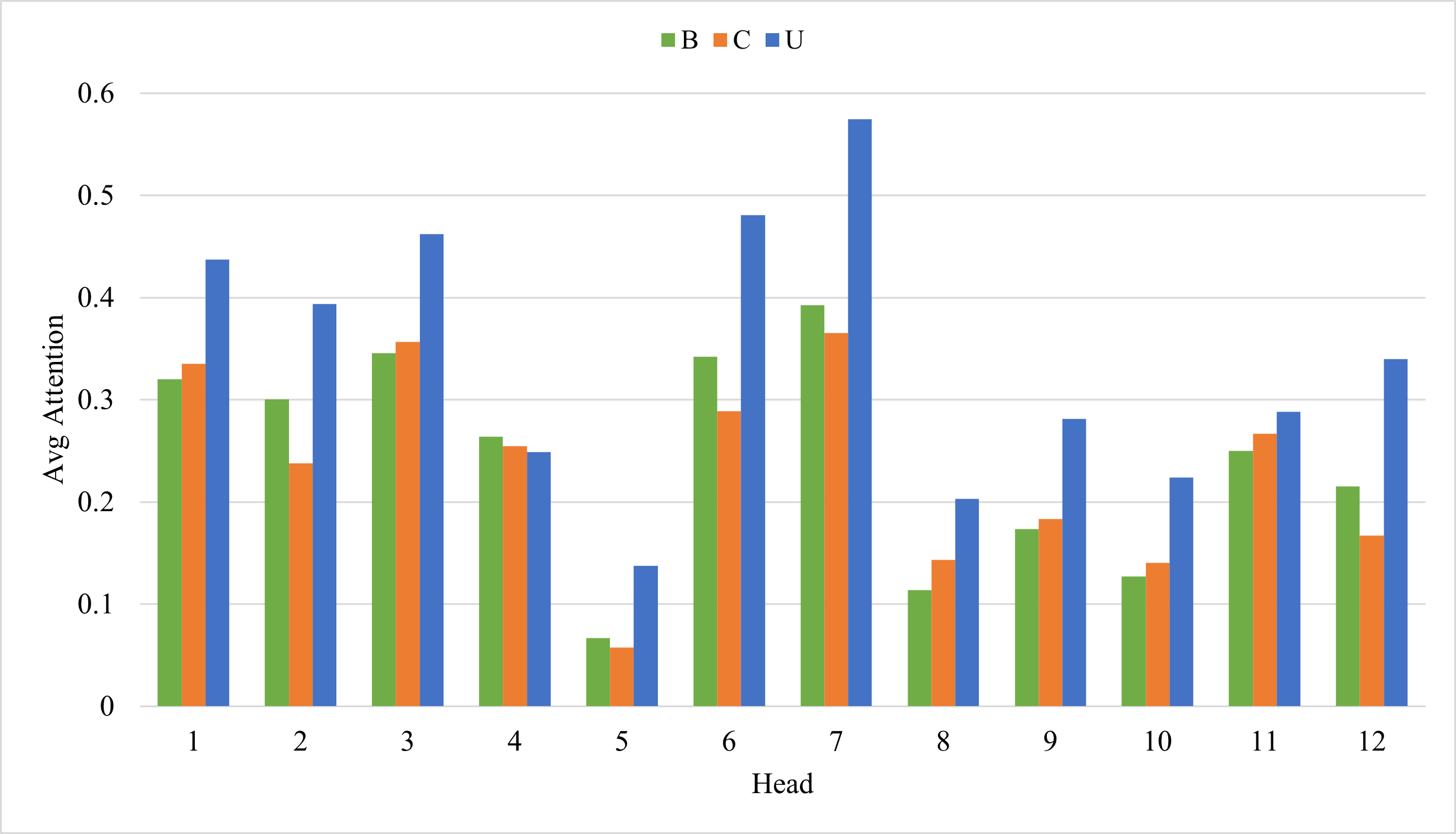}}
	\subfigure[CodeGen]{
		\label{fig:discussion-rq2.codegen}
		\includegraphics[width=\linewidth]{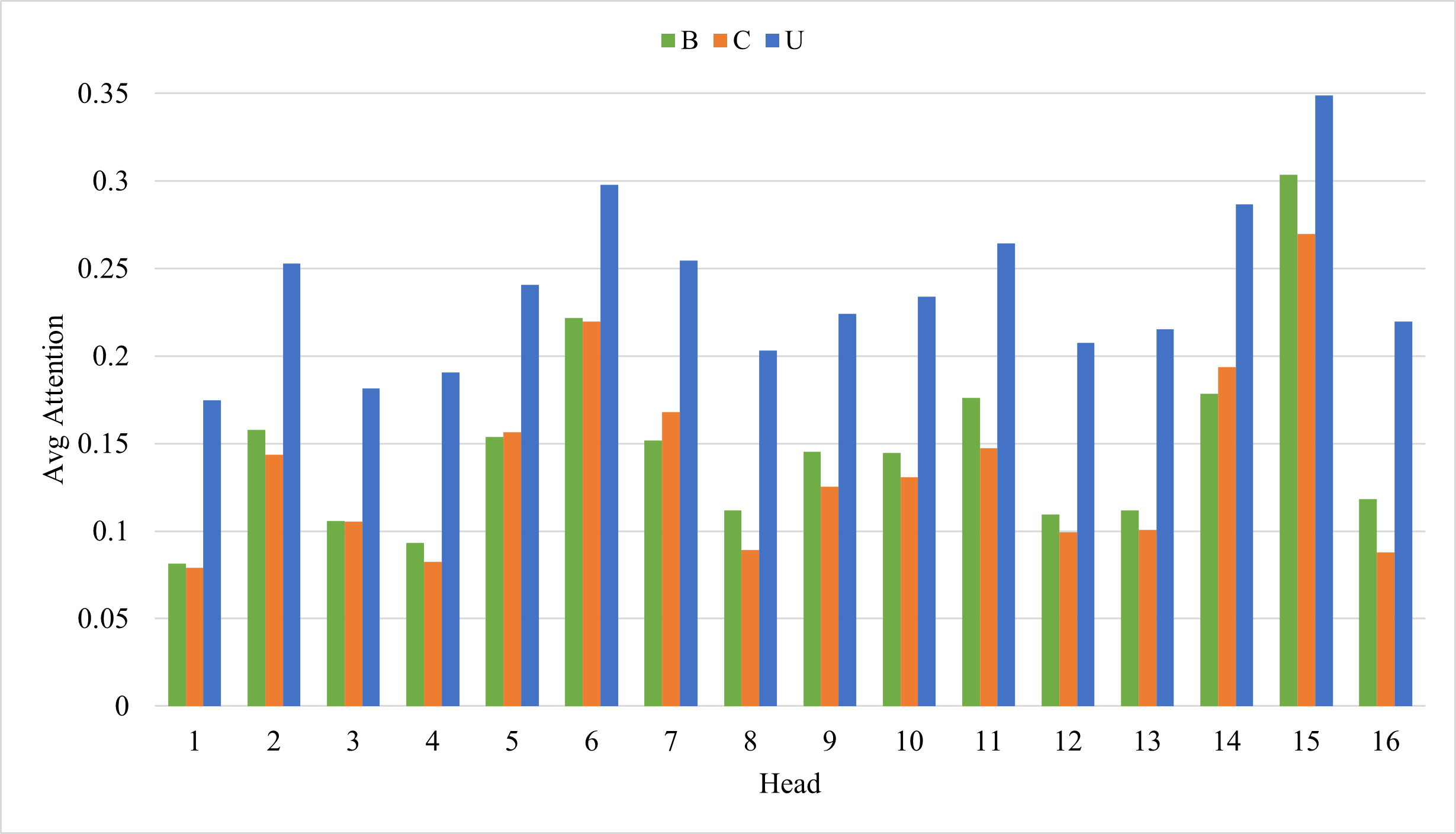}}
    \caption{Average attention scores of all attention heads on each input token itself using three different combinations of code and text of three different architecture code PTMs. Among them, subfigures (a), (b) and (c) represent encoder-only CodeBERT, encoder-decoder CodeT5 and decoder-only CodeGen respectively.}
   \label{fig:discussion-rq2}
\end{figure*}

\textbf{Discussion} We demonstrate that obtaining embeddings according to the way of inputting both code and text information during the pretraining of the PTMs results in lower-quality embeddings. In contrast, using unimodal input of code information and text information proves to be a competitive way to acquire higher-quality code embeddings. One possible reason is that employing bimodal input of both code and text, or concatenation, increases the input length. This leads to the PTMs not focusing on each individual code token itself when generating code embeddings, but rather dispersing attention to other tokens. 

To illustrate this phenomenon, we analyze the model's attention towards each input token itself during embedding generation. As our embeddings are derived from the output of the last layer's hidden states, we capture the attention of all attention heads in the final layer of the transformer. Figure \ref{fig:discussion-rq2} displays the average attention scores of all attention heads towards each input token itself when generating code embeddings using three different combinations of code and text for three different architecture code PTMs (namely, CodeBERT, CodeT5, and CodeGen). We observe that for unimodal input data, almost all attention heads of the three architecture PTMs show higher average attention scores towards each input token itself compared to bimodal input or concatenated input. This suggests that each token in unimodal input data garners increased attention from the model which means a richer embedded syntactic and semantic content, facilitating the final classifier in learning patterns specific to the corresponding classification task \cite{wan-discuss-attention, clark-discuss-attention}. On the other hand, with bimodal or concatenated input, due to the longer input length, the model pays attention not only to each token itself but also to more distant context tokens. However, these distant context tokens may not contribute semantically to the token itself that requires attention, resulting in relatively lower quality of the generated embeddings.

\textbf{\textit{No matter which architecture of the code PTMs is used, the quality of code embedding obtained by inputting data according to the way of inputting code and text information during pre-training tends to be subpar. This method cannot guarantee obtaining code embeddings with richer semantic information. On the other hand, the unimodal input of code information and text information proves to be a competitive approach for obtaining higher-quality code embeddings.}}

\section{Implications}
\label{sec: implication}
In this section, based on the experimental results of our above study, the following implications are drawn to guide researchers in further research.

\textbf{Implication 1) When researchers generate code embeddings for code snippets of classification tasks using code PTMs, it is advisable to aggregate vector representations of all code tokens to obtain the embeddings, rather than using the special token method commonly employed in the NLP field. } In this paper's experimental results of RQ1, we observed that classifiers constructed based on embeddings obtained through averaging the pooling of all code tokens outperform those built on embeddings obtained through special tokens in most cases. This trend holds on five PTMs of three different architectures across three distinct classification tasks. The performance improvement is evident across various metrics. This means that generating code embeddings should focus on all code tokens, rather than relying on empirical knowledge in the NLP domain. In addition, just a simple average-pooling of all code tokens can produce embeddings that contain richer semantic information. Therefore, focusing on all code tokens to obtain code embeddings is more in line with the data characteristics of SE classification task scenarios.

\textbf{Implication 2) Researchers should use decoder-only architecture code PTMs for generating embeddings of code snippets when dealing with classification tasks.} In the experimental results of RQ1, we also observed that when using special tokens to obtain embeddings, the embeddings generated by decoder architecture code PTMs do not surpass the classical encoder-only CodeBERT. This is true even for decoder-only architecture PTMs with larger parameters. However, by leveraging the aggregation of vector representations from all code tokens to obtain semantic embeddings, code PTMs of the decoder-only architecture can achieve semantic embeddings that are equally rich or even of higher quality compared to those obtained from encoder-only or encoder-decoder architectures. Given that decoder-only architecture PTMs have become the prevailing mainstream and are continually expanding in parameter size, we further recommend researchers to employ decoder-only architecture code PTMs for generating embeddings of code snippets.

\textbf{Implication 3) Researchers using code PTMs for generating code embeddings when processing classification tasks with code and text information should opt for unimodal input of code information and text information, followed by aggregation of these two parts to obtain higher-quality embeddings.} In the experimental results of RQ2 in this article, we observed that inputting data in the same way as code and text information during pre-training does not guarantee a code embedding with richer semantic information. However, the unimodal input method performs more prominently. This means that both code data and text data contain rich information. Connecting code and text will make the model focus too long and easily miss key information. The unimodal input enables the model to capture key information, and the generated embeddings contain pattern information of code and text respectively, which helps train the model for SE classification tasks. Therefore, when utilizing code PTMs to generate embeddings, we recommend unimodal input code and text, separately embedding and subsequently fusing them.

\section{Threats to Validity}
\label{sec: threats}
Although the experiments in this study demonstrate the sufficiency and effectiveness of our conclusions, the broader validity of these findings may still be subject to certain threats. This section discusses internal threats, construct threats and external threats to the validity of our work.
\subsection{Internal Validity}
Internal validity pertains to potential threats posed by objective factors to the conclusions of the experiments. The first concern is the dataset used for exploratory experiments. For both research questions in this paper, we utilized datasets that have been widely adopted by previous researchers. These datasets are publicly available and have undergone no tampering. Moreover, they stem from various projects and encompass different programming languages. The complexity and diversity of these datasets are sufficient to mitigate the related threats they might pose to the experimental conclusions. Furthermore, even though some research within their respective domains may rely solely on one or two evaluation metrics \cite{hoang-jit, dou-discuss}, we evaluate all classification tasks with three metrics. To ensure a comprehensive assessment of task performance and to eliminate any threat the choice of evaluation metric might pose to the experimental conclusions, we calculated the averages of Accuracy, F1, and MCC across 50 experiments for all tasks. Additionally, we performed statistical tests to guarantee a thorough and unbiased evaluation.
\subsection{Construct Validity}
Construct validity pertains to potential threats posed by the experimental setup and procedures on the conclusions of the experiments. Since this paper primarily focuses on code embeddings, for classification tasks, we employ the same fully connected layer for classification across different embeddings. The performance of this classifier is used to gauge the quality of code embeddings. To mitigate any potential impact of data preprocessing on task performance, we follow common practices in the task domain \cite{kanade-cubert, hoang-jit, liu-ccrep}. This involves using source code data and text data as inputs to obtain the corresponding embeddings, without the need for intricate preprocessing, as our goal is not to enhance the performance of downstream tasks. Additionally, we ensure that each task inputs code data and text data of the same length into different models, to guarantee that the richness of semantic information in the embeddings will not be affected by variations in input information volume. Finally, in the process of obtaining code embeddings using each pre-trained model, we make every effort to utilize previously open-sourced code to ensure correctness. However, the possibility of manual errors still exists. We make all data and code used in the experiments publicly available and deposit them into an open-source repository to ensure the replicability of the experiments and promote future research. We hope to engage researchers in related fields and continue to integrate more in-depth studies.
\subsection{External Validity}
External validity pertains to factors affecting the generalizability of our conclusions. The experimental findings in this paper are derived from using five code PTMs that generate code embeddings across four SE classification tasks. However, we still cannot guarantee that these experimental conclusions can be extrapolated to all SE classification tasks and other code PTMs. In terms of generating embeddings, we utilize code PTMs covering all three different architectures, making the conclusions valid for different architectures. For code PTMs, we employ popular models belonging to these three different architectures: CodeBERT, CodeT5, PLBART, CodeGPT, and CodeGen. These five models generate embeddings of varying scales. However, with the rising popularity of PTMs, future research should also consider more types of embeddings, such as GraphCodeBERT \cite{guo-graphcodebert} and LLaMA \cite{touvron-llama}. The former employs richer code structural information to generate embeddings, while the latter is an extremely large-scale universal language model. Regarding SE tasks, this paper selects four classification tasks that involve different input content and programming languages. Nevertheless, exploring more SE classification tasks would further enrich our conclusions. Furthermore, recent studies indicate that different computing devices may affect task performance \cite{tang-gpu}. All experiments in this paper are conducted on a consumer-grade NVIDIA GTX 2080Ti GPU. However, we believe that conducting experiments on the same computing device should yield conclusions similar to ours. Since we do not engage in any adjustments to the embeddings throughout the entire experiment, different computing devices only affect the speed of the experiment, not the experimental conclusions.

\section{Conclusion}
In this paper, we conduct a systematic study on how to generate higher-quality code embeddings on four code understanding tasks using a total of five code PTMs with three different architectures. Specifically, we studied whether two commonly used methods for obtaining code embeddings by researchers in the current SE field are effective. Our experimental results show that the method commonly used by researchers to obtain code embeddings through special tokens is not suitable for code classification tasks, and it is also difficult to obtain high-quality embeddings by inputting code information and text information according to the combination of code and text during pre-training of the pre-trained model. We recommend that SE researchers and practitioners pay attention to all code tokens of the input data when generating embeddings for code snippets and unimodally input code information and text information to obtain semantically richer embeddings for downstream tasks. Our findings provide guidance for the SE field on how to use code PTMs to generate better embeddings as well as clues for future research on reusing code PTMs.


\bibliographystyle{ieeetr}
\bibliography{ref.bib}

\end{document}